\documentclass[aps,twocolumn,amsmath,amsfonts,nofootinbib,preprintnumbers,superscriptaddress,eqsecnum,secnumarabic,floatfix]{revtex4-1}
\usepackage{graphicx,slashed}
\usepackage[utf8]{inputenc}
\usepackage{amsmath}
\usepackage{amssymb}
\usepackage{tikz}
\usepackage[caption = false]{subfig}
\newcommand{\be}{\begin{equation}}
\newcommand{\ee}{\end{equation}}
\def\bsp#1\esp{\begin{split}#1\end{split}}

\begin{document}
\title{
Dark matter from dark photons:  \\
\vspace{0.5mm}
a taxonomy of dark matter production}

\author{Thomas~Hambye}
\email{thambye@ulb.ac.be}
\affiliation{Service de Physique Th\'eorique,
Universit\'e Libre de Bruxelles\\
Bld du Triomphe CP225, 1050 Brussels, Belgium}
\affiliation{Instituto de F\'isica Te\'orica, IFT-UAM/CSIC, U.A.M., Cantoblanco, 28049 Madrid, Spain}

\author{Michel~H.G.~Tytgat}
\email{mtytgat@ulb.ac.be}
\affiliation{Service de Physique Th\'eorique,
Universit\'e Libre de Bruxelles\\
Bld du Triomphe CP225, 1050 Brussels, Belgium}

\author{J\'{e}r\^{o}me~Vandecasteele}
\email{jvdecast@ulb.ac.be}
\affiliation{Service de Physique Th\'eorique,
Universit\'e Libre de Bruxelles\\
Bld du Triomphe CP225, 1050 Brussels, Belgium}

\author{Laurent Vanderheyden}
\email{lavdheyd@ulb.ac.be}
\affiliation{Service de Physique Th\'eorique,
Universit\'e Libre de Bruxelles\\
Bld du Triomphe CP225, 1050 Brussels, Belgium}

\date{\today}

\preprint{ULB-TH/19-07}
\begin{abstract}
We analyse how dark matter (DM) can be produced in the early universe, working in the framework of a hidden sector charged under a U(1)' gauge symmetry and interacting with the Standard Model through kinetic mixing.
Depending on the masses of the dark matter particle and of the dark photon, as well as on the hidden U(1)' gauge coupling and the kinetic mixing parameter, we classify  all the distinct regimes along which the observed dark matter relic density can be accounted for. We find that  9 regimes are potentially operative  to produce the DM particles and this along 5 distinct dynamical mechanisms. Among these, 4 regimes 
{are new and correspond to regimes in which the DM particles are produced by on-shell dark photons. One of them proceeds along a new dynamical mechanism, which we dub sequential freeze-in.}
We argue that such regimes and the associated dynamical mechanisms are characteristic of DM models for which, on top of the Standard Model and the dark sector, there are  other massive, but relatively light particles --- akin to the dark photon --- that interact both with the SM and the DM sectors. 
\end{abstract}

\maketitle

% =====================================================================
\section{Introduction}

As the nature of the dark matter (DM) remains a mystery, it is possible that DM is a particle that belongs to a whole new hidden sector. This hidden sector may be coupled through the Standard Model (visible sector) through a few {possible} portals \cite{Patt:2006fw}.  
The question we study further in the present work is how to account, in generic terms, for the abundance of a dark matter (DM) particle produced through portals. For the case of the kinetic mixing portal with a massless dark photon, this has been addressed in much details \cite{Chu:2011be}. That work also includes the case of dark matter creation through the Higgs portal. In the present work, we consider  the possible impact of a finite dark photon mass. More generically, our study applies to DM production in models in which a relatively light particle couples both to the DM and to SM particles.

The existence of a massive dark photon,
 associated to a hidden U(1)' gauge interaction, has been the object of many investigations, both theoretically and experimentally, see e.g. the reviews \cite{Jaeckel:2010ni,Essig:2013lka,Alexander:2016aln}. This possibility is well-motivated, very rich phenomenologically and, moreover, is directly related to the DM problem. Indeed, a gauge symmetry is a most natural way to stabilize a particle \cite{Foot:2014uba,Ackerman:mha,Feng:2008mu,Feng:2009mn,Hambye:2010zb}. Specifically, we consider the  following simple and popular model
\begin{eqnarray}
\label{eq:model}
\mathcal{L}&\supset & -\frac{1}{4}B'^{\mu\nu}B'_{\mu\nu} -\frac{\hat\epsilon}{2}B^{\mu\nu}B'_{\mu\nu} +\frac{1}{2}m_{\gamma'}^{2}B'^{\mu}B'_{\mu}\nonumber\\
&& +\; i\bar{\chi}\slashed{D}\chi - m_{\rm DM}\bar{\chi}\chi + \ldots .
\end{eqnarray}
Here $\chi$ is a Dirac fermion, singlet under the SM gauge group but charged under U(1)' and will be our  DM candidate. Its covariant derivative is $D_\mu = \partial_\mu+ie' B'_\mu$, where $B'_\mu$  and $e'$ are the U(1)' gauge field and coupling.  This  dark, or hidden, sector is coupled to the Standard Model (SM) sector through 
the so-called kinetic mixing portal term, which mixes the dark gauge  field to the SM hypercharge one with a mixing parameter $\hat{\epsilon}$ \cite{Holdom:1985ag} . This Lagrangian thus involves 4 new parameters,
$m_{\rm DM}$, $\hat{\epsilon}$, $e'$ and what will turn out to be the mass of the dark photon particle, $m_{\gamma'}$. Equivalently, we will make use of the hidden sector fine structure constant $\alpha'\equiv e'^2/4\pi$ and of the combination $\kappa =\hat  \epsilon\, \cos\theta_W\,e'/e $, which is the millicharge of the DM in the limit $m_{\gamma^\prime} \rightarrow 0$. The dark photon mass can arise either through the St\"uckelberg \cite{Stueckelberg:1900zz} or through the  Brout-Englert-Higgs  mechanism \cite{Englert:1964et,Higgs:1964pj}. In the latter case, there are necessarily several other parameters and degrees of freedom associated  to U(1)' breaking (the dots in Eq.~(\ref{eq:model})). In the sequel, we will assume that these extra ingredients may be neglected. However, we will briefly discuss their possible impact in Appendix \ref{sec:BEH}.

In the case of a massless dark photon, the problem of accounting for the DM abundance  through the kinetic mixing portal has been studied in  \cite{Chu:2011be}. It has been shown in that work that the observed relic density could be reached along 4 distinct dynamical mechanisms depending on the values of the parameters of the model. These dynamical mechanisms are:  freeze-in (regime Ia in the sequel) \cite{McDonald:2001vt,Hall:2009bx,Chu:2011be,Bernal:2017kxu},  reannihilation  (IIIa) \cite{Hall:2009bx,Chu:2011be},   secluded freeze-out (IVa)\footnote{By secluded freeze-out, we refer here to a scenario in which DM particles freeze-out  in the dark sector  \cite{Pospelov:2007mp,Pospelov:2008zw} but with a temperature $T'$ different from the 
temperature of the SM visible sector, as in \cite{Chu:2011be, Feng:2009mn}.} and, finally, the standard text-book thermal freeze-out mechanism (Va and Vb). Altogether, these 4 mechanisms lead to 5 different regimes through which the DM abundance can be reached in the case of a massless dark photon. Indeed, the standard freeze-out of the dark matter particles could occur either through annihilation into SM particles (Vb regime) or into the dark photons themselves (Va regime). We will thus distinguish production mechanisms (classified using Roman numerals) and regimes (distinguished using the Latin letter a or b, as in the case of freeze-out). 

\begin{center}
\begin{figure}[h!]
\includegraphics[scale=1.0]{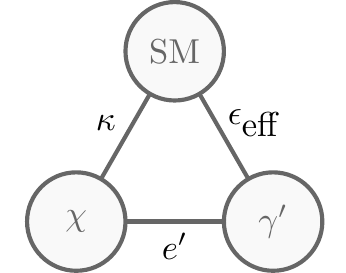}
\caption{The  3 sectors (blobs) and their 3 connections (lines). In the limit of massless dark photon, (or more generally, as $\epsilon_{\rm eff} \rightarrow 0$, see text)  the connection between the SM and dark photons blob is absent.}
\label{fig::Triangle}
\end{figure}
\end{center}

The aim of the present work is to revisit this classification of DM production mechanisms, taking into account the possible effects of the mass of the dark photon. By considering the dark photon model above, the generic underlying structure we will be considering is that of a system composed of 3 distinct particle sectors, together with 3 possible connections between the sectors. This structure is depicted in Fig.~\ref{fig::Triangle}, again with a focus on the kinetic mixing setup and its ingredients.  So, the 3 sectors (depicted as blobs) consist of the SM, the dark matter particle $\chi$ and the massive dark photon $\gamma'$, while the connections between the sectors (depicted as lines) are parameterized by the mixing $\hat \epsilon$ (or more precisely $\epsilon_{\rm eff}$, which will be defined below), the hidden fine structure constant $\alpha'$ and the "millicharge" $\kappa$. 
{Considering such a structure, we will point} out the existence {of  4 new regimes (noted Ib, II, IIIa and IVa),  including one along a new dynamical mechanism (II), making} in total potentially 9 distinct ways to reach the observed {DM abundance}.

\begin{figure}[h!]
\subfloat[Ia]{\includegraphics[height=0.28\linewidth]{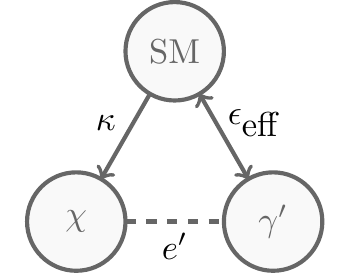}}
\subfloat[Ib]{\includegraphics[height=0.28\linewidth]{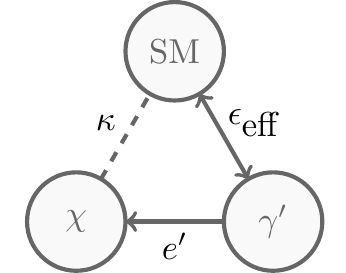}}\\
\subfloat[II]{\includegraphics[height=0.28\linewidth]{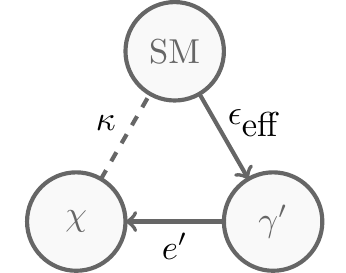}}\\
\subfloat[IIIa \& IVa]{\includegraphics[height=0.28\linewidth]{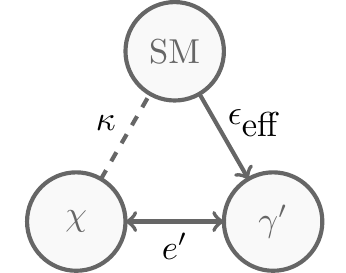}\label{fig::7TriangleIIIa}}
\subfloat[IIIb \& IVb]{\includegraphics[height=0.28\linewidth]{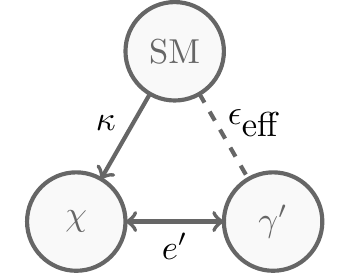}\label{fig::7TriangleIIIb}}\\
\subfloat[Va]{\includegraphics[height=0.28\linewidth]{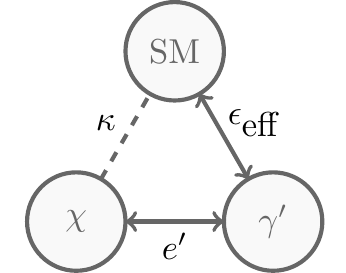}}
\subfloat[Vb]{\includegraphics[height=0.28\linewidth]{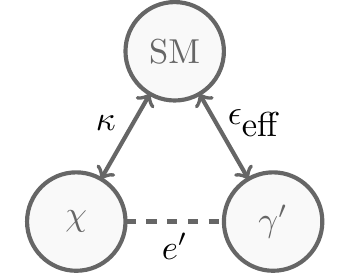}}\\
 \label{fig::7Triangle}
 \caption{The 9 possible DM production regimes in the dark photon scenario. A double sided arrow means that the two corresponding sectors have reached chemical equilibrium; 
a single sided arrow indicates slow out-of-equilibrium production of one sector by the other one; a dashed line corresponds to a subdominant interaction between the sectors. Regimes Ia and Ib, II, IIIa and IIIb, IVa and IVb, Va and Vb are associated to 5 distinct mechanisms to produce the DM abundance:  the freeze-in (I), sequential freeze-in (II), reannihilation (III), secluded freeze-out (IV) and freeze-out (V) mechanisms respectively (see Sections \ref{subsec:FI} and \ref{subsec:SFI}). Notice that the diagrams are identical for the reannihilation and secluded freeze-out mechanisms. Note that the SM to $\gamma'$ connection is parametrized by  $\epsilon_{\rm eff}$ when we take into account the thermal corrections. Without such corrections, it is parametrized by $\epsilon$.}
\end{figure}

As we will explain in the following sections, the 9 production regimes can be read from this figure by considering the different ways through which the hidden sector particles can be created and whether they are in thermal equilibrium with each others and/or with the SM sector. This is shown in Fig.~\ref{fig::7Triangle}. There, one sees the 4 new regimes that emerge in the case of massive dark photon (and so are absent in the  massless limit --- see later) and noted Ib, II, IIIa and (a priori, see however below) IVa. As the solid arrow lines between the sectors mean to suggest, they all correspond to regimes in which DM is created from the SM sector via the production of real, on mass-shell dark photons.  All these regimes follow each others along a characteristic pattern in the parameter space, which in the sequel we call "the phase diagram" and that can be seen in Figs.~\ref{fig::DogDiagram1} and~\ref{fig::DogDiagram2} for two illustrative hidden sector mass setups. Most interestingly, some of these new regimes allow for DM production for values of $\kappa$ that are even smaller than in the case of standard freeze-in  (regime Ia). 
As we will see, whether a specific regime is actually relevant will depend not only on the connections parameters, but also on the DM and dark photon masses.

Our work is organized as follows. In Section \ref{sec:sectors} we will present how the various particles in the kinetic portal model of (\ref{eq:model}) can interact, focusing on the case of a massive dark photon. Next, in Section \ref{sec:phases} we will derive and discuss the parameter space or phase diagram in which we show, as a function of the parameters, how the different regimes that lead to the observed DM relic density are related with each others. 
We discuss each regime in details but put a particular emphasis on the new regimes. For the sake of our classification and to simplify our discussion, in that section we will be putting aside some complications that arise due to the peculiar nature of the kinetic mixing portal. In particular, we {neglect there the} impact of thermal effects on the propagation and production of dark photons. These will be considered in Section \ref{sec:ThermalEffects}, emphasizing their impact on the different production regimes we  found. Next we will discuss briefly, in Section \ref{sec:constraints}, the other constraints which hold on this model. Finally, in Section \ref{sec::gen}, we will elaborate on the generality of the phase diagram we have obtained, discussing the possible effect of other degrees of freedom and then draw our conclusions. The Appendix \ref{app:1} contains a technical summary based on the existing literature on thermal effects on dark photons production and progagation. The case in which the mass of the dark photon arise through the Brout-Englert-Higgs mechanism is considered in Appendix \ref{sec:BEH}. 

%%%%%%%%%%%%%%%%%%%%%%%%%%%%%%%%%%%%%%%%%%%%%%%
\section{The three sectors and their connections}
\label{sec:sectors}
 
We focus on the model {of} Eq.~(\ref{eq:model}) and aim at studying the abundance of the new particles, here the dark matter, {made of particle $\chi$ and $\bar \chi$}, and the dark photon $\gamma'$. To determine the relevant processes, we need to establish the coupling of dark photons to the SM sector. The procedure is standard and consists  first to have canonical kinetic terms for the gauge fields; we repeat the argument here for the sake of clarity. To do so, we exploit the fact that the values of  $\epsilon$ that are relevant for DM production  will turn out to be always small numbers, so we can treat  the effects of mixing as a perturbation. From Eq.~(\ref{eq:model}), making the non-orthogonal transformation
\begin{equation}
B'^\mu = \tilde B'^\mu - \hat \epsilon B^\mu
\end{equation}
leads to 
\begin{eqnarray}
\mathcal{L}&\rightarrow & -\frac{1}{4}\tilde B'^{\mu\nu}\tilde B'_{\mu\nu} +\frac{1}{2}m_{\gamma'}^{2}\tilde B'^{\mu}\tilde B'_{\mu}\nonumber - \hat \epsilon\,  m_{\gamma'}^{2} B^{\mu}\tilde B'_{\mu} \\
&&  - e' \bar{\chi}\gamma^\mu\chi  (\tilde B'_\mu  - \hat \epsilon B_\mu),  \label{eq:Lagrangian2}
\end{eqnarray}
to leading order in $\hat\epsilon$, where we only show the terms relevant for our argument. Clearly, as the SM particles only couple to the SM gauge fields, in particular the neutral ones, $B^\mu = \cos\theta_W A_0^\mu - \sin\theta_W Z_0^\mu$ and $W^3_\mu= \sin\theta_W A^\mu_0 + \cos\theta_W Z^\mu_0$, in the limit $m_{\gamma'}=0$, they do not mix with the dark photon $\left(\equiv \tilde B'_\mu\right)$. Here $\theta_W$ is the usual Weinberg angle; the subscript $0$ is to insist on the fact that they are the usual SM fields. The SM particles do not interact with the dark photon while the $\chi$ particles are coupled to the SM photon $A^\mu_0$ with millicharge $\kappa = \epsilon e'/e  $, where $\epsilon =\hat \epsilon\cos\theta_W$. They are also coupled vectorially to the $Z^\mu_0$ boson, with coupling $- e' \epsilon \tan\theta_W$. Thus, if $m_{\gamma'} = 0$, the only possibility to create dark photons is through the production of $\chi$ particles \cite{Foot:1994bx,Chu:2011be}.

If, on the other hand, $m_{\gamma'} \neq 0$, 
 dark photons can be created directly by SM particles through the mixing mass term
\begin{equation}
\label{eq:mixing_mass}
- \hat \epsilon\,  m_{\gamma'}^{2} \tilde B'_{\mu}B^{\mu}\equiv - \epsilon\, m_{\gamma'}^2 \tilde B'_\mu (A_0^\mu - \tan\theta_W Z_0^\mu),
\end{equation}
 in Eq.~(\ref{eq:Lagrangian2}). Like in the description of neutrino oscillations, we refer to the basis of Eq.~(\ref{eq:Lagrangian2}) as the interaction eigenstates basis. It is interesting to go forward and diagonalize the mass terms. We will only consider cases in which $m_{B'} \ll m_Z$, and so again we do the diagonalization to leading order in $\epsilon \ll 1$. Performing the orthogonal transformation 
\begin{equation}
A'^\mu = \tilde B'^\mu - \epsilon A_0^\mu \quad\mbox{\rm and}\quad A^\mu =  A_0^\mu + \epsilon \tilde B'^\mu 
\end{equation}
transforms Eq.~(\ref{eq:mixing_mass}) into 
\begin{equation}
\label{eq:mixing_mass2}
-\epsilon\, \tan\theta_W\, m_{\gamma'}^2 A'_\mu \ Z_0^\mu,
\end{equation}
while $m_{A'} = m_{\gamma'}$, with $m_{Z_0}$ like in the SM. 
Now, it is easy to see that this mixing term introduces a mass splitting between the $A'^\mu$ and $Z_0^\mu$ field that is ${\cal O}(\epsilon^2)$ and thus can be neglected. So the dark photon field $A'_\mu$ has indeed a mass
\begin{eqnarray} 
m_{A'} & = &m_{\gamma'},
\end{eqnarray}
 while the field $A^\mu$ is  massless. This eigenmass basis makes clear that the particles $\chi$ couple only to the massive dark photon and to the SM $Z$ boson, and not to the massless photon, a fact that is well-known,
 \begin{equation}
 e' \bar{\chi}\gamma^\mu\chi  (\tilde B'_\mu  - \hat \epsilon B_\mu) \equiv  e' \bar{\chi}\gamma^\mu\chi  (A'_\mu + \epsilon \tan\theta_W Z_\mu) 
 \end{equation}
 (we drop the $0$ subscript on the $Z$ field from now on as $m_{Z_0} \equiv m_Z$). Also, the electrically charged SM particles, like the electron, couple to the dark photon with coupling
 \begin{equation}
 e \bar \psi\gamma^\mu \psi A_{0\mu} \equiv e \bar \psi\gamma^\mu \psi (A_\mu + \epsilon A'_\mu).
 \end{equation}

\begin{figure}[h!]
\subfloat[]{\includegraphics[height=0.25\linewidth]{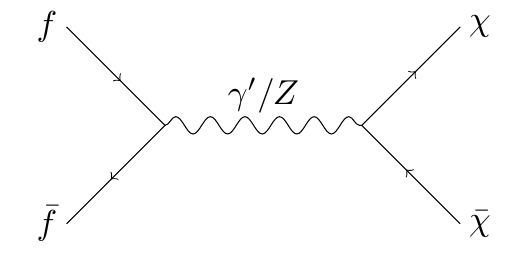}}
\subfloat[]{\includegraphics[height=0.25\linewidth]{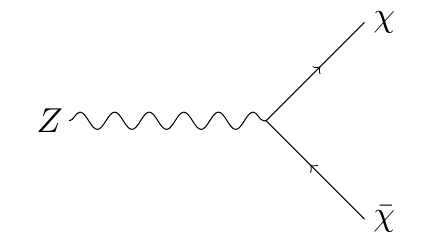}}\\
\subfloat[]{\includegraphics[height=0.25\linewidth]{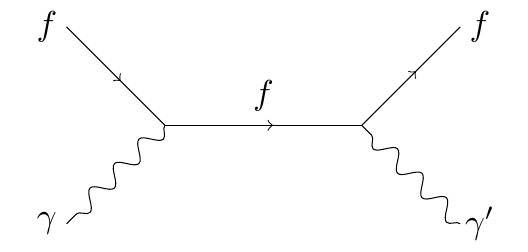}}
\subfloat[]{\includegraphics[height=0.25\linewidth]{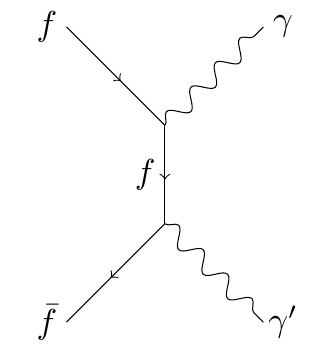}}\\
\subfloat[]{\includegraphics[height=0.25\linewidth]{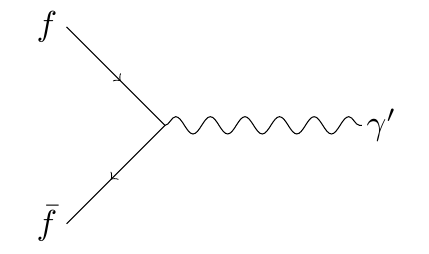}}
\subfloat[]{\includegraphics[height=0.25\linewidth]{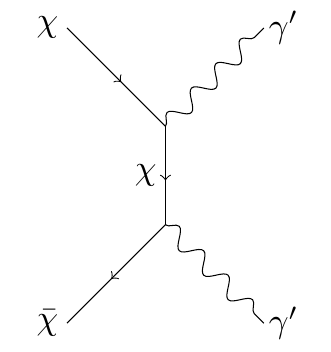}}\\
\caption{Feynman diagrams for all the connecting processes between the three populations.}
\label{fig::Feynman}
\end{figure}

In the model just introduced we must thus consider the existence of three distincts populations (or sectors, or reservoirs) of particles in the early stages of the universe. The main reservoir consists of the SM particles; they will be assumed to be in thermal equilibrium at a temperature $T$. 
The hidden sector consists on one side of the $\chi$ particles and their antiparticles (DM) and, on the other side, of the dark photons. These two populations do not have to be { necessarily}  in thermal equilibrium with each others or with the SM sector. Their respective abundance depends then on how the 3 reservoirs are connected with each others. 
{The dominant processes directly}  connecting the SM and DM populations are pair annihilation of  SM particles (or $Z$ boson decay for some mass range) into DM pairs \cite{Chu:2011be}. The strength of these processes is essentially set by a single parameter (the "millicharge"), 
\begin{equation}
\label{eq:kappa}
\kappa \equiv  \epsilon \sqrt{\alpha '/\alpha}
\end{equation}
The processes connecting the SM reservoir and the dark photon population are on the other hand determined by the 
$\epsilon$ parameter;
these {\em a priori} include Bremsstrahlung $e^\pm  e^\pm \rightarrow e^\pm e^\pm \gamma'$ (which may be neglected), Compton scattering $e^\pm  \gamma \rightarrow e^\pm \gamma'$, pair annihilation $e^+ e^-  \rightarrow \gamma \gamma'$, and coalescence $e^+ e^-  \rightarrow  \gamma'$. They involve distinct powers of $\alpha$ but also distinct temperature dependence and so are dominant in different mass and temperature regimes. As for the processes connecting the DM and dark photon populations, they are set by $\alpha'$ and are dominated (at lowest order in $\alpha'$) by $\gamma'\,\gamma'\leftrightarrow \chi \bar \chi$. 
The Feynman diagrams for all these connecting processes are shown in Fig.~\ref{fig::Feynman}.
This structure  leads to the diagram of Fig.~\ref{fig::Triangle}, where the blobs represent schematically the reservoirs and the lines the possible connections. In the massless case
the triangle of Fig.~\ref{fig::Triangle} has one connecting edge less since there is no direct connection between the hidden photon and  SM populations. {This, in brief, explains why there are more DM production regimes in the massive case than in the massless case.}

%%%%%%%%%%%%%%%%%%%%%%%%%%%%%%%%%
\section{Phase diagram of dark matter production}
\label{sec:phases}

We now turn to the systematic discussion of the dependence of the DM relic abundance on the connector parameters, starting from an empty reservoir in the dark sector,
passing through all the regimes, up to a dark sector that is fully in thermal equilibrium with the SM.  For the sake of clarity and generality,  we will not include in this section any of the thermal effects that are specific to the dark photon setup. Instead, thermal effects on dark photon production, which can be very important in some cases as we shall see, will be determined and discussed in Section \ref{sec:ThermalEffects}. Also, in this section and the sequel, we will rely heavily on the results obtained in Ref.~\cite{Chu:2011be}, to which we refer for more in-depth discussions of some of the production regimes, including explicit expressions for the relevant cross-sections (given in that reference for a massless dark photon). 

\begin{center}
\begin{figure}[h!]
\includegraphics[scale=0.6]{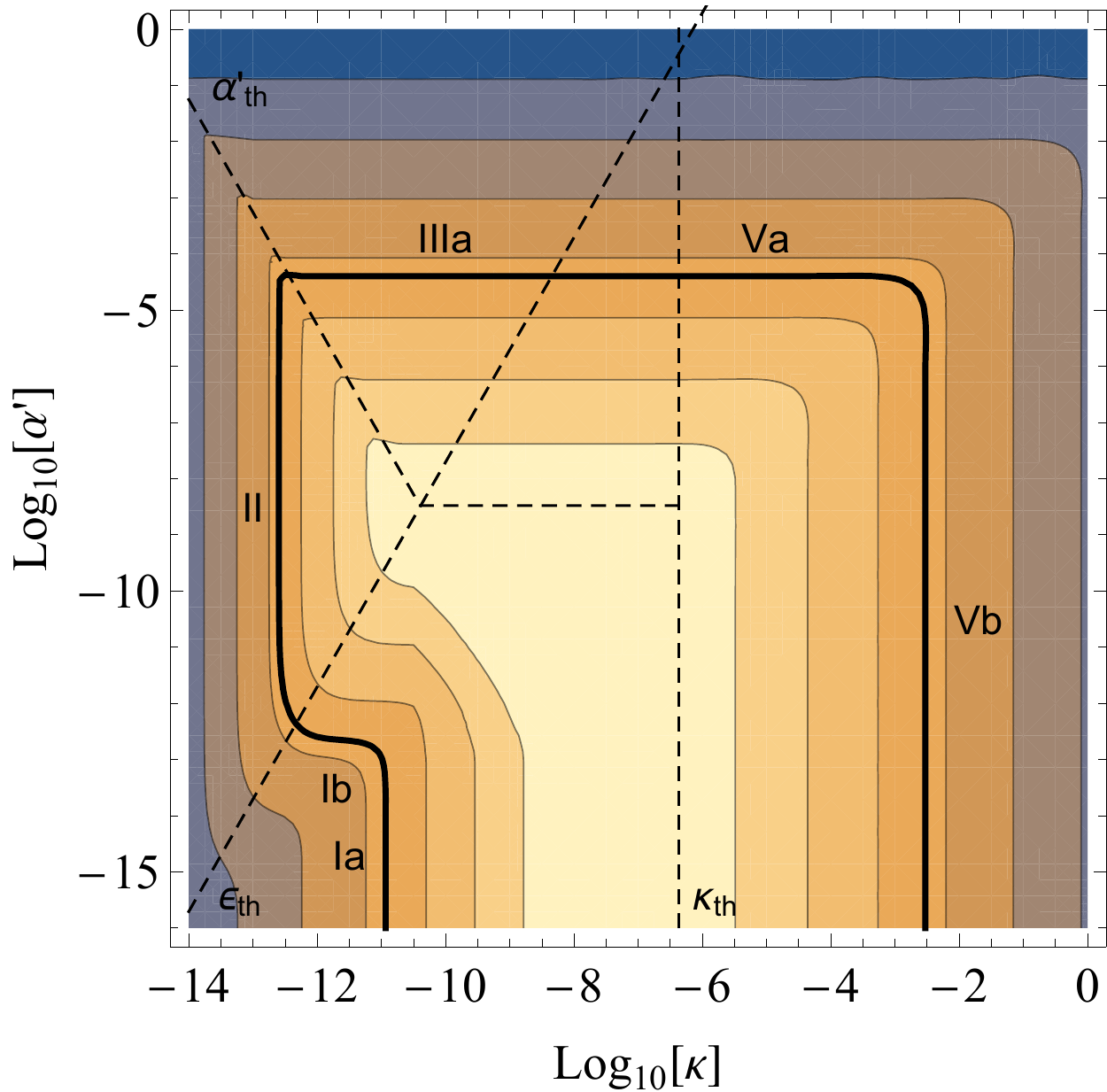}
\caption{\rm DM relic density obtained as a function of $\kappa$ and $\alpha'$ for $m_{\rm DM}=3$ GeV and $m_{\gamma '}=1$ GeV. This diagram displays the various different production regimes which can lead to the observed relic density without taking thermal effects into account. For this particular choice of masses one obtains 6 different production regimes along 4 dynamical ways: freeze-in (Ia $\&$ Ib), sequential freeze-in (II), reannihilation (IIIa) and freeze-out (Va $\&$ Vb).}\label{fig::DogDiagram1}
\end{figure}
\end{center}

\begin{center}
\begin{figure}[h!]
\includegraphics[scale=0.6]{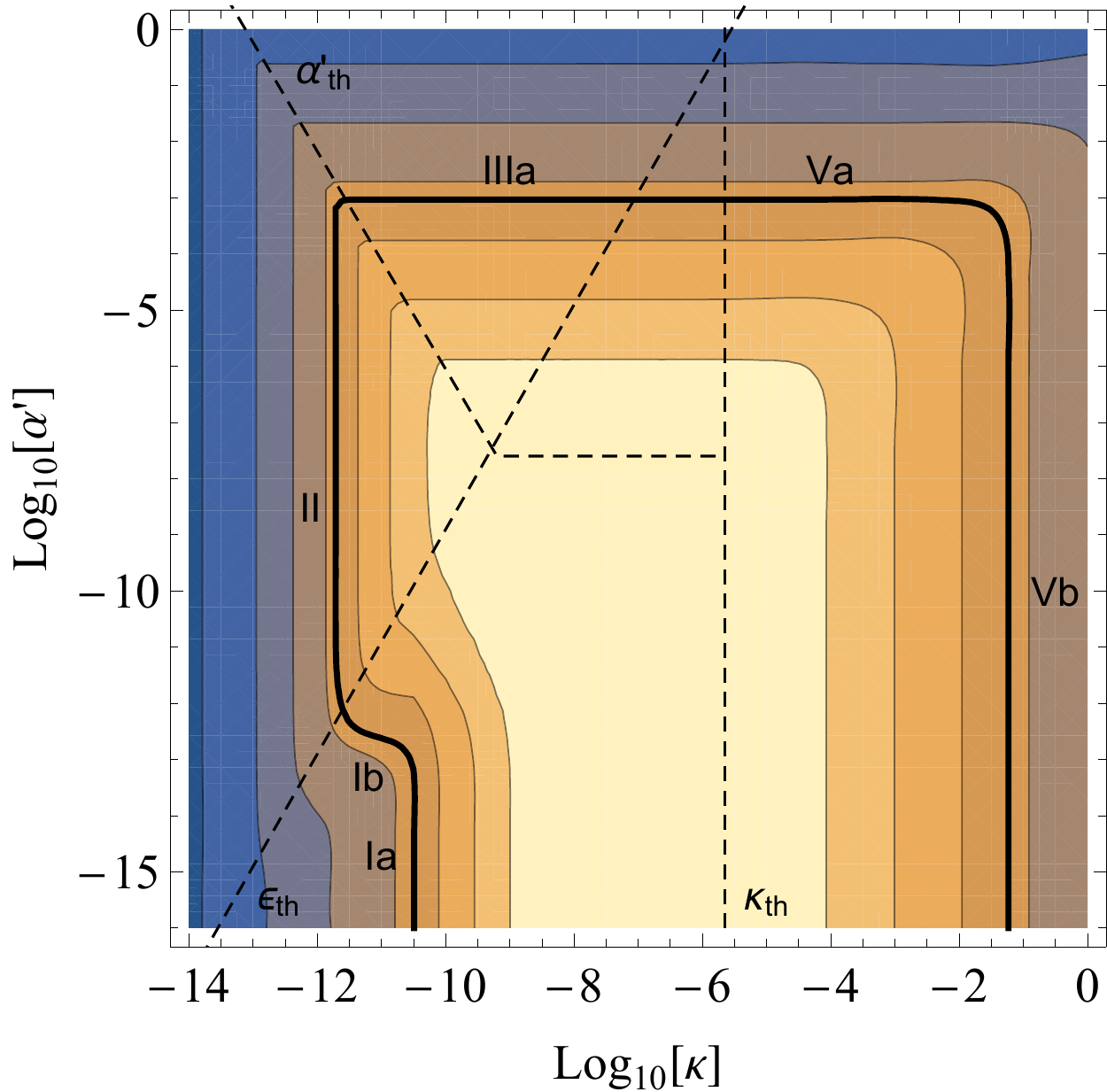}
\caption{Same as Fig.~\ref{fig::DogDiagram1} but with $m_{\rm DM}=100\text{ GeV}$ and $m_{\gamma'}=10\text{ GeV}$.}
\label{fig::DogDiagram2}
\end{figure}
\end{center}

The DM abundance depends on the DM and dark photon masses and two among the three connector parameters $\kappa$, $\alpha '$ and $\epsilon_{\rm eff}$, since only two are independent, see Eq.~(\ref{eq:kappa}). Thus, for a given set of  $\chi$ and $\gamma '$ masses, one can for instance give the relic abundance as contour lines in the $\kappa-\alpha '$ plane. This representation leads to a ``phase diagram" that displays the various regimes. Note importantly that for all the discussion below we will always assume that $m_{\gamma'} < m_{\rm DM}$ so that DM is always  kinematically free to annihilate into dark photons. The opposite case  $m_{\gamma'}> m_{\rm DM}$ leads to a different {(and simpler) phenomenology} that we will not discuss here.
The phase diagram  comes out from integrating a set of Boltzmann equations that determine the evolution  of the DM and $\gamma'$ yields as function of the visible sector temperature.
For the case of  a massless dark photon, 
the contour lines of constant DM relic density in the phase diagram have roughly the shape of a rectangle (dubbed the ``mesa'' in \cite{Chu:2011be}), displaying 5 regimes, along 4 dynamical production mechanisms. These 4  mechanisms are freeze-in (Ia in the classification of Fig.~\ref{fig::7Triangle}), reannihilation (IIIb), secluded freeze-out (IVb) and ordinary freeze-out (with freeze-out either  to dark photons, Va, or SM particles, Vb, leading  all together to 5 distinct regimes).
Starting with an empty hidden sector, the DM abundance is reached through the following sequence of regimes
\begin{equation}
{\rm Ia \rightarrow IIIb \rightarrow IVb \rightarrow Va \rightarrow Vb} \quad (m_{\gamma'} = 0).
\end{equation}
{In particular, the freeze-in regime, Ia, proceeds through slow $\kappa$ driven $\rm SM \rightarrow DM$ processes and} leads to a vertical line at small $\kappa$ in the "mesa" structure of the phase diagram, see \cite{Chu:2011be}.
When the dark photon mass matters, this simple vertical line structure does not hold anymore and instead one has a more complicated structure for small $\kappa$.
Actually, one can distinguish 4 more regimes for values of $\kappa$ that are small enough so that the $\rm SM\rightarrow DM$ processes do not thermalize. This is 
illustrated by the phase diagram depicted in Fig.~\ref{fig::DogDiagram1} which we get
  for one example set of masses, $m_{\rm DM}=3$ GeV and $m_{\gamma '}=1$ GeV. Much heavier DM candidates are of course possible, as shown in Fig.~\ref{fig::DogDiagram2}, for $m_{\rm DM}=100$ GeV and $m_{\gamma '}=10$ GeV.
More to the point, the phase diagram, which shows contour lines of constant DM relic density, has a distinct shape: the "mesa" has an extension toward much smaller value of the $\kappa$ parameter, suggesting the shape of a ``mooring bollard''. These features, and the corresponding new regimes, are, as we shall see, due to the possibility of dark photon production of DM.

{The most generic sequence of regimes appearing along the mooring bollard pattern is the one appearing in Figs.~\ref{fig::DogDiagram1} and~\ref{fig::DogDiagram2}, that is to say
\begin{equation}
{\rm Ia \rightarrow Ib \rightarrow II \rightarrow IIIa \rightarrow Va \rightarrow Vb} \quad ({m_{\gamma'} \neq 0}).
\end{equation}
This structure, on top of the regimes already existing in the massless case (Ia, Va and Vb), involve 3 new regimes: Ib, II and IIIa. The fourth new regime, IVa, as well as the 2 other regimes already existing in the massless case (IIIb and IVb), 
can appear for other choices of the parameter and more generally for other models. The four new regimes arise from the extra SM production of dark photons.}
In all cases the mooring bollard shape of the phase diagram is a generic signature.

The general set of Boltzmann equations that determine  the evolution of the DM and $\gamma'$ abundances ($Y_{\rm DM,\gamma'} = n_{\rm DM,\gamma'}/s$, with $s$ the entropy density and $n_{\rm DM} = n_\chi + n_{\bar \chi}$) as a function of time, and so that leads to the phase diagram, takes the form\footnote{Here and in subsequent Boltzmann equations, we have included  factors of $1/2$, typical of Dirac DM particles, into the definitions of the cross-sections \cite{Gondolo:1990dk}. }
\begin{eqnarray}
zHs\frac{dY_{\rm DM}}{dz}&=&\,\langle \sigma_{\rm DM \rightarrow SM}\, v\rangle \,\left[({n_{\rm DM}^{\rm eq}})^2-n^2_{\rm DM} \right]\nonumber\\
&&\hspace{-0.27cm}+\,\langle \Gamma^D_{\rm DM \rightarrow SM} \rangle\frac{n_{Z}^{\rm eq}}{(n_{\rm DM}^{\rm eq})^{2}} \left[{(n_{\rm DM}^{\rm eq}})^2-n^2_{\rm DM} \right]\label{eq:YDM}\nonumber\\
&&\hspace{-0.27cm}+\,\langle \sigma_{\gamma'\rightarrow \rm DM} v\rangle n^2_{\gamma'}-\,\langle \sigma_{\rm DM\rightarrow \gamma '} v\rangle n^2_{\rm DM}, \\
zHs\frac{dY_{\gamma '}}{dz}&=&\langle \sigma_{\gamma ' \rightarrow \rm SM}\, v\rangle\, n_{\rm SM}^{\rm eq}\,[{n_{\gamma '}^{\rm eq}}-n_{\gamma'}]
\nonumber\\
&&\hspace{-0.27cm}+ \,\langle \sigma_{\rm DM \rightarrow \gamma '} v\rangle n^2_{\rm DM}-\,\langle \sigma_{\rm \gamma'\rightarrow DM} v\rangle n^2_{\gamma'},
\label{eq:YDP}
\end{eqnarray}
where $H$ is the Hubble parameter and $z = m_{\rm DM}/T$, with $T$ the temperature of the visible sector. 
The quantity $\Gamma^D$  refers to the $Z$ decay rate into a pair of DM particles. In writing these Boltzmann equations, we assumed that the dark photon is lighter than the dark matter (excluding for instance the channel $\gamma' \rightarrow \chi \bar \chi$). 
A sum over the different $\rm SM\leftrightarrow DM$ and $\rm SM\leftrightarrow \gamma'$ channels is implicit  everywhere in these equations. In the sequel, so as to avoid cluttering of the equations, we will regroup the scattering and decay terms involving SM particles into
\begin{equation}
\gamma_{\rm SM \leftrightarrow DM}^{\rm eq} =  \langle \sigma_{\rm DM \rightarrow SM}\, v\rangle ({n_{\rm DM}^{\rm eq}})^2+ {\langle \Gamma^D_{\rm SM \rightarrow DM} \rangle} {n_{\rm SM}^{\rm eq}}
\end{equation}
and
\begin{equation}
\gamma_{\rm SM\leftrightarrow \gamma'}^{\rm eq} = \langle \sigma_{\rm \gamma '\rightarrow SM}\,v\rangle n_{\rm SM}^{\rm eq} n_{\gamma'}^{\rm eq}.
\end{equation}

Although the Boltzmann equations contain many terms, for most production regimes, only one or two of these terms are relevant.
As we will see too, these equations are not sufficient to correctly determine the amount of DM produced in the reannihilation regimes, which are characterized by a hidden and visible sectors with distinct temperatures: in these cases, one also needs 
to evaluate the energy that has been transferred from the SM to the hidden sector particles.

{To understand} the distinction between the various production regimes, it is useful to start by delimiting the regions of parameter space depending on whether the various connecting processes lead, or not, to thermalization. 
To determine whether the DM  particles thermalize with the SM thermal bath, we take the simple criteria
\begin{align}
\left. \frac{\Gamma_{\rm SM\leftrightarrow DM}}{H}\right |_{T\sim m_{\rm DM}}\gtrsim 1 \, ,\label{eq:DMThermSM}
\end{align}
with $\Gamma_{\rm SM\leftrightarrow DM}=\gamma_{SM \leftrightarrow DM}^{\rm eq}/n_{\rm DM}^{\rm eq}(z)$. This leads to the following condition on the millicharge parameter, 
\begin{align}
\kappa\gtrsim \kappa_{\rm th}\equiv 3.8\times 10^{-7}\left(\frac{m_{\rm DM}}{\text{GeV}}\right)^{1/2},
\label{kappath}
\end{align}
which is depicted by a vertical line in Figs.~\ref{fig::DogDiagram1} and \ref{fig::DogDiagram2}.
As for the thermalization of the dark photons with the SM thermal bath, what matters typically is if  thermalization (which leads to $E_{\gamma'}\sim T$) has occurred by the time when the DM number freezes, $T\sim m_{\rm DM}$. Thus similarly to the condition above, we use
\begin{align}
\left. \frac{\Gamma_{\rm SM\leftrightarrow \gamma'}}{H}\right|_{T\simeq m_{\rm DM}}\gtrsim 1 \, ,\label{eq:DPThermSM}
\end{align}
as the condition for the thermalization of the dark photons with the SM, 
which numerically translates in a condition on the (in principle effective, but we remind the reader that we neglect the thermal effect for the time being) mixing parameter, 
\begin{align}
\epsilon \gtrsim \epsilon_{\rm th}\equiv 4.1\times 10^{-8}\left(\frac{m_{\rm DM}}{\text{GeV}}\right)^{1/2},
\label{epsilonth}
\end{align}
with $\Gamma_{\rm SM\leftrightarrow \gamma'}=\gamma_{\rm SM \leftrightarrow \gamma'}^{\rm eq}/n_{\gamma'}^{\rm eq}(z)$. As $\kappa = \epsilon \sqrt{\alpha'/\alpha}$, this condition corresponds to a  diagonal dashed line in the phase diagram, see Figs.~\ref{fig::DogDiagram1} and \ref{fig::DogDiagram2}.
As for the thermalization between the dark photons and DM particles themselves, it depends on the scenario considered. 
We  now discuss each of the new regimes, using the phase diagram of Fig.~\ref{fig::DogDiagram1} as reference. 

%%%%%%%%%%%%%%%%%%%%%%%%%%%%%%%%%%%%%
\subsection{Freeze-in:  regimes Ia and Ib}
\label{subsec:FI}

The Ia freeze-in regime corresponds to production of DM particles directly from the SM particles, through out-of-equilibrium processes $\rm SM \rightarrow DM$  parameterized by $\kappa$. In this regime, $\alpha'$ is assumed to be too small for the processes $\rm \gamma' \rightarrow DM$ to play any role. Thus, in Eq.~(\ref{eq:YDM}) only the $\rm SM\rightarrow DM$ term  is relevant for producing the DM.
\begin{equation}
\hspace{-0.5cm}\hbox{\underline{Regime Ia}}\,:\qquad zHs\frac{dY_{\rm DM}}{dz} \approx \gamma_{\rm SM \leftrightarrow DM}^{\rm eq}(z).
\end{equation}
For fixed DM mass, the production depends only on the parameter $\kappa$ so the observed relic abundance is given by a vertical line in the phase diagram, corresponding to the lower left corner in Fig.~\ref{fig::DogDiagram1}. 

Note that for small value of $\alpha'$, and with the value of $\kappa$ that this freeze-in regime requires $\kappa \sim 10^{-10}$, values of $\epsilon$ lie above the critical value $\epsilon_{\rm th}$ for thermalization of dark photons Eq.~(\ref{epsilonth}). Thus, unlike for the $\rm SM\rightarrow DM$ freeze-in scenario in the massless dark photon case, in this instance the dark photons thermalize with the SM sector, forming a single thermal bath characterized by the temperature $T$.
However, this thermalization does not change anything to the dynamics of this freeze-in regime as a function of $T$, since the DM particles are created dominantly by the SM particles. The  only effect is  that the thermalized $\gamma'$ modify the number of relativistic degrees of freedom and thus the Hubble expansion rate. The modification is of order $g_\gamma'/g^{\rm SM}_\star=2/g^{\rm SM}_\star \sim 10^{-2}$, a small effect that we neglect 
in Fig.~\ref{fig::DogDiagram1}. 

\begin{figure}
\includegraphics[width=8cm]{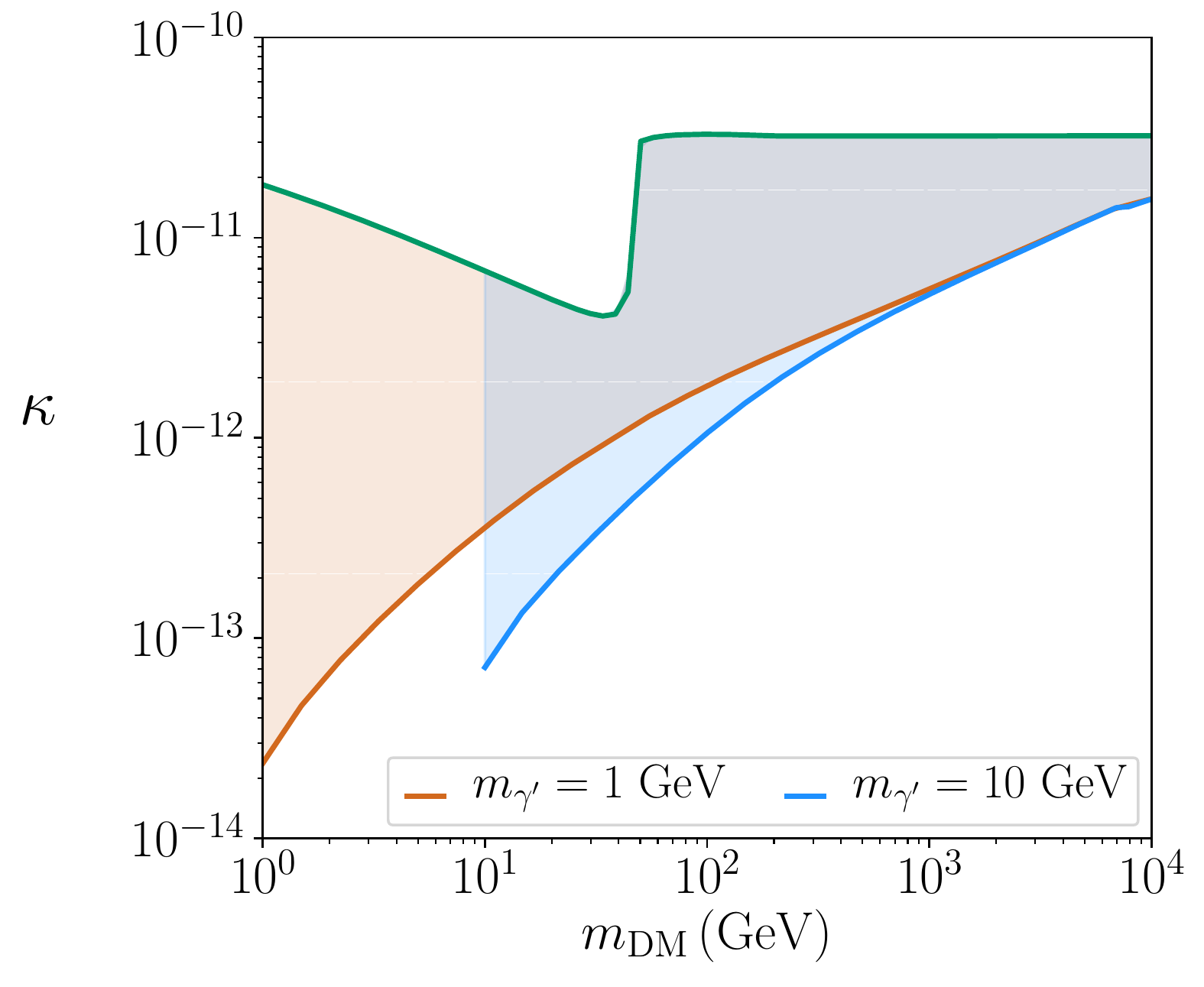}
\caption{{Values of $\kappa$ needed to account for the observed relic density, as a function of $m_{DM}$, for the standard freeze-in (regime Ia, green line) and for the sequential freeze-in (regime II, orange line for $m_{\gamma '}=1$ GeV and blue line for $m_{\gamma '}=10$ GeV) in the $\kappa-m_{\rm DM}$. The shaded regions thus correspond to regime Ib, that is freeze-in from thermalized dark photons. We consider $m_{\gamma'} < m_{\rm DM}$. }}
\label{fig:FIAndDFIPS}
\end{figure}

In Fig.~\ref{fig:FIAndDFIPS} we show as a function of $m_{\rm DM}$  the $\kappa$ required  to reach the observed relic abundance in the Ia regime, which we note $\kappa_{\rm Ia}$. Effectively, as the thermalized dark photons play a negligible role in the production of the DM particles, this curve is essentially (modulo a slight dependence on the dark photon mass in the dark photon propagator) the same as in the massless dark photon case \cite{Chu:2011be}.  The dependence of $\kappa_{\rm Ia}$ on $m_{\rm DM}$ is complicated by the fact that the dominant production channels depend also on $m_{\rm DM}$, but otherwise the number of DM particles created through $\rm SM_i \,SM_i\rightarrow DM\,DM$ is simply related to equilibrium quantities evaluated at a temperature $T\ll {\rm max}\{m_i,m_{\rm DM}\}$, 
\begin{equation}
Y_{\rm DM}(z)=\sum_i c_i\,\frac{({n_i^{\rm eq}})^2\,\langle \sigma_{\rm SM_i\rightarrow DM} v\rangle}{Hs}\Big|_{T={\rm max}\{T,m_i,m_{\rm DM}\}},
\label{nDMFIa}
\end{equation}
where the $c_i$ are coefficients of order unity and that depends on the channel considered \cite{Chu:2011be}. 

Going up along the vertical line of the Ia regime depicted in Fig.~\ref{fig::DogDiagram1}, the value of $\alpha'$ increases while that of $\epsilon$ decreases. Thus, at some point, $\epsilon$ become smaller than $\epsilon_{\rm th}$ (given in Fig.~\ref{fig::DogDiagram1}  by the {diagonal dashed} line). However, before this could happen and while the dark photons are still in thermal equilibrium with the SM,  $\alpha'$ becomes large enough  for the thermal dark photons to {sizeably} pair produce DM particles. So, in this case, DM becomes dominantly produced through $\rm \gamma'\rightarrow DM$ freeze-in instead of $\rm SM\rightarrow DM$ freeze-in. Clearly, in this case, a smaller value of $\kappa$ is required to avoid over-production of DM particles and as a result the abundance depends only on $\alpha'$, giving rise to the horizontal  line depicted in Fig.~\ref{fig::DogDiagram1}. This ``freeze-in from dark photons'' regime is denoted as Ib in this figure.\footnote{This  regime has been briefly discussed in \cite{Kane:2015qea}, where it is dubbed "inverse annihilation", and in \cite{Klasen:2013ypa}, in a model with a scalar singlet that mixes with the Higgs.}
For this regime the only relevant term in the Boltzmann equation is the one producing DM from dark photon:
\begin{equation}
\hspace{-1.5cm}\hbox{\underline{Regime Ib}}\,:\qquad zHs\frac{dY_{\rm DM}}{dz}\approx \gamma^{\rm eq}_{\rm \gamma' \leftrightarrow DM}\label{BoltzEqIb},
\end{equation}
with $\gamma^{\rm eq}_{\rm \gamma' \leftrightarrow DM} = \langle \sigma_{\rm \gamma'\leftrightarrow DM} v\rangle (n^{\rm eq}_{\gamma'})^2$
so that
the DM yield is simply given by
\begin{equation}
Y_{\rm DM}(z) =c_{\gamma'}\,\frac{({n_{\gamma'}^{\rm eq}})^2(z)\,\langle \sigma_{\rm \gamma' \rightarrow DM} v\rangle}{Hs}\Big|_{T = Max[T,m_{\rm DM}]},
\label{nDMFIb}
\end{equation}
with $c_{\gamma'} = {\cal O}(1)$. 
In the limit $m_{\gamma'} \ll m_{\rm DM}$, we found that the value of $\alpha'$ required to create the right amount of DM particles out-of-equilibrium from the $\gamma'$ is
\begin{align}
\alpha'_{\rm Ib}&=2.5\times 10^{-13} \left(\sqrt{g_\ast\left(m_{\rm DM}\right)}g_{\ast,S}\left(m_{\rm DM}\right)\right)^{1/2},
\label{alphaFI}
\end{align}
where the dependence on the in equilibrium SM degrees of freedom stems from the $Hs$ factor in the denominator {of Eq.~(\ref{nDMFIb}).
As stated above, the} value of $\kappa$ in the Ib regime must be below that of Ia, $\kappa_{\rm Ia}$, so as to avoid over-production of DM. However, if it becomes much smaller than $\kappa_{\rm Ia}$, we enter in yet another regime, which we now discuss.

%%%%%%%%%%%%%%%%%%%%%%%%%%%%%%%
\subsection{Sequential freeze-in: regime II}
\label{subsec:SFI}

As we move along the Ib regime line towards smaller values of $\kappa$ in Fig.~\ref{fig::DogDiagram1}, the value of $\epsilon$ decreases. Thus at some point $\epsilon$ becomes smaller than $\epsilon_{\rm th}$, i.e.~the dark photons no longer thermalize with the SM. At this point, clearly Eq.~(\ref{BoltzEqIb}) does not apply since $n_{\gamma'}$ is not anymore equal to its equilibrium value $n^{\rm eq}_{\gamma'}$.
Thus the system enters a new regime in which none of the processes induced by $\kappa$, $\epsilon$ or $\alpha'$ have ever been in thermal equilibrium. Also, the $\rm SM\rightarrow DM$ freeze-in processes induced by $\kappa$ are clearly too slow to be efficient since $\kappa\ll\kappa_{\rm Ia}$. However, it turns out that the slow out-of-equilibrium production of dark photons from the SM (which is controlled by $\epsilon$) followed by the slow out-of-equilibrium production of DM by these un-thermalized dark photons can 
produce enough DM. We thus have a chain of successive freeze-in processes, something we dub "sequential freeze-in", see regime II in Fig.~\ref{fig::DogDiagram1}.
{As  $\rm DM \rightarrow \gamma'$ and} $\rm \gamma'\rightarrow SM$ processes have a negligible impact in this regime, they drop from the Boltzmann equations which take the form
\begin{eqnarray}
\hspace{-2cm}\hbox{\underline{Regime II}}\,:\quad&&\nonumber\\
&&\,\,\, zHs\frac{dY_{\gamma '}}{dz}\approx \gamma ^{\rm eq}_{SM\leftrightarrow\gamma '},\label{BoltzEqII1}\\
&&zHs\frac{dY_{\rm DM}}{dz}\approx \gamma_{\gamma '\leftrightarrow DM},
\label{BoltzEqII2}
\end{eqnarray}
where $\gamma_{\gamma '\leftrightarrow DM}=\langle \sigma_{\gamma'\leftrightarrow DM} v\rangle (n_{\gamma'})^2$.
With respect to the equilibrium reaction density, this reaction density is suppressed by a $(n_{\gamma'}/n_{\gamma'}^{\rm eq}(z))^2$ factor, with $n_{\gamma'}$ determined by the first Boltzmann equation, Eq.~(\ref{BoltzEqII1}). 
As the number of dark photons produced by SM particles is proportional to $\epsilon^2$, the number of DM particles, which is proportional to $n_{\gamma'}^2
 \,\langle \sigma_{\gamma'\rightarrow DM}\,v\rangle$, is $\propto \epsilon^4{\alpha'}^2 \sim \kappa^4$. Thus, in the sequential freeze-in regime, the relic density depends only on $\kappa$ (as in Ia freeze-in regime, albeit with a smaller value of $\kappa < \kappa_{\rm Ia}$) and corresponds  to a vertical line in the phase diagram, as seen in Fig.~\ref{fig::DogDiagram1}. 
 
The line separating regime Ib and II is the line for which $\epsilon=\epsilon_{\rm th}$. Consequently, the value of $\kappa$ required to reach the DM abundance in regime II  is set  by setting $\epsilon=\epsilon_{\rm th}$ while taking $\alpha' = \alpha'_{\rm Ib}$  for the regime Ib, Eq.~(\ref{alphaFI}). For example, for the masses considered in Fig.~\ref{fig::DogDiagram1}
this gives 
\begin{align}
\kappa_{\rm II}=\epsilon_{\rm th} \sqrt{\frac{\alpha '_{\rm Ib}}{\alpha}}\simeq 10^{-13} \ll \kappa_{\rm Ia},\label{eq: KappaSFI}
\end{align}
in good agreement with the numerical value shown in this figure. 
This shows that along  sequential freeze-in it is possible to have a SM to DM connection ({\em i.e.} millicharge $\kappa$) which may be orders of magnitude weaker than along the ordinary freeze-in regime: for the example of Fig.~\ref{fig::DogDiagram1} $\kappa_{\rm II}$  is 3 orders of magnitude smaller than $\kappa_{\rm Ia}$.

We give in Fig.~\ref{fig:FIAndDFIPS} the value $\kappa_{\rm II}$ as a function of $m_{\rm DM}$ corresponding to the {observed DM relic density}. This is aimed at illustrating the fact that sequential freeze-in requires a smaller value of $\kappa$. Thermal effects {below will somehow} change the details of the picture, but not this overall conclusion, which we deem to be general for model with relatively light mediators. 
Fig.~\ref{fig:FIAndDFIPS} also  shows that, for large values {of $m_{\rm DM}/m_{\gamma'}$, $\kappa_{\rm II}$ tends} to $\kappa_{\rm Ia}$,  meaning that the only relevant mechanism to produce DM in this case is regime Ia in this region of parameter space. This merging of regime II (and thus of the intermediate regime Ib too) with the ordinary Ia freeze-in regime for large values {of $m_{\rm DM}/m_{\gamma'}$ is due to the fact that more massive DM candidates are produced at higher temperatures and thus at earlier times}. If the dark photons have less time to thermalize with the SM, they will need a comparatively larger mixing to achieve it.
Thus, the turning from phase Ib to II occurs for a {larger} value of $\kappa$, making $\kappa_{\rm II}$ to get closer to $\kappa_{\rm Ia}$.\footnote{When we will take into account the finite temperature corrections, we will see that the merging of regime $\kappa_{\rm II}$ and $\kappa_{\rm Ia}$ occurs for {somewhat} smaller value {of $m_{\rm DM}/m_{\gamma'}$, see} Fig.~\ref{fig:FIAndDFIPSThermal} and Section \ref{sec:ThermalEffects}.}

%%%%%%%%%%%%%%%%%%%%%%%%%%%%%%%%%%%%%%%% 
\subsection{Reannihilation: regimes IIIa and IIIb}
\label{sec:reannihilation}

Moving in Fig.~\ref{fig::DogDiagram1} along regime II towards larger values of $\alpha'$, clearly, at some point $\alpha'$ becomes large enough for the
$\rm DM$ and the dark photon to reach thermal equilibrium so that the system enters a new regime called reannihilation. As explained at length in Ref.~\cite{Chu:2011be}, this regime holds if two conditions are fulfilled. First, it requires that thermalization of the dark sector occurs without thermalization with the SM sector. Consequently, the  system is composed of two thermal baths (the SM  and the hidden sector) characterized by two different temperatures, $T$ and $T'$ with $T > T'$. Second, this regime requires that the slow out-of-equilibrium production of dark sector particles from the SM ones is still efficient at the time of freeze-out of the DM into hidden sector particles (which takes place at a temperature  $T' \leq m_{\rm DM}$). In this case, DM freezes later than in the case of a standard freeze-out in the hidden sector (i.e. a ``secluded freeze-out", see below) and  undergoes a period of reannihilation. 

 In the present case, reannihilation could be achieved in two distinct ways, depending on whether the hidden sector is populated through $\rm SM\rightarrow \gamma'$ (controlled by $\epsilon$) or through $\rm SM\rightarrow DM$ (respectively $\kappa$) slow processes. These regimes are noted IIIa and IIIb respectively in Figs.~\ref{fig::7TriangleIIIa}, \ref{fig::7TriangleIIIb} and \ref{fig::DogDiagram1}.
Now, in practice, if one moves towards larger values of $\alpha'$ along the sequential freeze-in regime (regime II) line in Fig.~\ref{fig::DogDiagram1}, the system enters into regime IIIa and not regime IIIb.
This simply stems from the fact that, with a value of $\kappa$ as small as the one which holds for regime II and which is smaller than for the ordinary $\rm SM\rightarrow DM$ freeze-in Ia regime,
 DM particles can not be {sizeably} produced through $\rm SM\rightarrow DM$ processes. Consequently, we begin with regime IIIa and discuss next regime IIIb.

The value of $\alpha'$ for which the transition from II to IIIa regimes occurs corresponds to the minimum value of $\alpha'$ for which the $\rm DM \leftrightarrow \gamma'$ processes thermalize
before the number of DM particles freezes. {This is determined by requesting that the rate for $\rm \gamma'\rightarrow DM$  is larger than the Hubble rate when the out-of-equilibrium $\rm SM\rightarrow \gamma'$ source stops creating dark photons with an energy large enough for the  $\rm \gamma'\rightarrow DM$ process to proceed, that is when $T\sim m_{DM}$,
\begin{align}
\left. \frac{\left\langle\sigma_{\rm \gamma '\rightarrow DM}v \right\rangle n_{\gamma '}}{H}\right|_{T\simeq m_{\rm DM}}\gtrsim 1. \label{eq::DMDPTherm} 
\end{align}
In this equation, the number density of dark photons that have been created out-of-equilibirium from the SM, $n_{\gamma'}$, can be related  to their number density if they where in thermal equilibrium with the SM thermal bath,  $n_{\gamma '}^{\rm eq} $,  {multiplying it by the square of the
ratio between $\epsilon$ and  the value  $\epsilon_{\rm th}$ that leads to thermalization, given in Eq.~(\ref{epsilonth}), so that} $n_{\gamma '}=\left(\epsilon/\epsilon_{\rm th}\right)^2 n_{\gamma '}^{\rm eq} (T)$. Thus,} the value of $\alpha'_{\rm IIIa}$ at the II to IIIa transition {follows from  the one} required for the $\rm DM\leftrightarrow \gamma'$ processes to thermalize if the dark photon was in thermal equilibrium with the SM, as
\begin{align}
\alpha'_{\rm IIIa}=\frac{\epsilon_{\rm th}}{\epsilon}\alpha'_{\rm th}.\label{condII-IIIa}
\end{align}
{This gives}
\begin{align}
\alpha'_{\rm th} \approx 2\times 10^{-9}\left(\frac{m_{\rm DM}}{\text{GeV}}\right)^{1/2}.
\end{align}

Once the dark photons  and the DM particles have thermalized, the abundance of $\gamma'$ and DM particles
are governed by the following set of Boltzmann equations:
\begin{eqnarray}
\hspace{-1cm}\hbox{\underline{Regime IIIa}}\,:\,\,\,
zHs\frac{dY_{\gamma '}}{dz}&=&\gamma^{\rm eq}_{\rm SM \leftrightarrow \gamma '} (z)\quad\quad\quad
\nonumber\\
&&\hspace{-2.5cm}\quad\quad-\,\gamma^{\rm eq}_{\rm \gamma '\leftrightarrow DM}  (z')   \left[1-\left(\frac{Y_{\rm DM}}{Y_{\rm DM}^{\rm eq}\left(z'\right)}\right)^2\right],\label{eq:DRARP}\\
zHs\frac{dY_{\rm DM}}{dz}
&=&\gamma^{\rm eq}_{\rm \gamma '\leftrightarrow DM}  (z')  
\nonumber\\
&& \times \left[1-\left(\frac{Y_{\rm DM}}{Y_{\rm DM}^{\rm eq}\left(z'\right)}\right)^2\right].\label{eq:DRADM}
\end{eqnarray}
The $\rm \gamma'\leftrightarrow DM$ terms are responsible for the thermalization between the $\gamma'$ and DM particles. 
They account for the fact that the $\gamma'$ number density is the one at equilibrium taken at temperature $T'$, i.e.~$Y_{\gamma'}^{\rm eq}(z')$, so that the corresponding reaction rate { is $\gamma^{\rm eq}_{\rm \gamma'\leftrightarrow DM}(z')$. The first} term of the first equation describes the slow production of dark photons from the SM.

The above Boltzmann equations can be solved provided we know $T'$ as a function of $T$. This is determined from integrating the Boltzmann equation for SM to hidden sector energy transfer \cite{Chu:2011be}
\begin{eqnarray}
zH\frac{d\rho '}{dz}+4H(\rho '+p')&=&
\,({n_{\rm SM}^{\rm eq}}(z))^2 \langle \sigma_{\rm SM\rightarrow DM} \,v\,\Delta E \rangle\nonumber\\
&&\hspace{-0.5cm}+({n_{\rm SM}^{\rm eq}}(z))^2 \langle \sigma_{\rm SM\rightarrow \gamma'}\, v\,\Delta E \rangle.\quad\quad
\end{eqnarray}
In reannihilation regime IIIa, the second term dominates the process of energy transfer.
Plugging the equation of state $p'(\rho')$ into this equation allows to determine $\rho'$, which in turns gives $T'$ as solution of
\begin{equation}
\rho '=\rho_{\gamma '}^{\rm eq}\left(z'\right)+\rho_{\rm DM}^{\rm eq}\left(z'\right)\,.
\end{equation}
For instance, when $T'\lesssim m_{\rm DM}$, the equation of state is $p'=\left(\rho '-m_{\rm DM} Y_{\text{\rm DM}}s\right)/3$, and $\rho_{\rm DM}\left(z'\right)=\rho_{\rm DM}^{\rm eq}\left(z\right)Y_{\rm DM}/Y_{\rm DM}^{\rm eq}(z)$.   

The DM production dynamics and the final amount of DM particles obtained from these Boltzmann equations along the reannihilation regime have been discussed in Ref.~\cite{Chu:2011be}.
The final amount of DM it gives, $\Omega_{\rm DM}$, approximately scales as $\log (\langle \sigma_{\rm eff} v\rangle)/\langle \sigma_{\rm \gamma'\rightarrow DM} v\rangle$ (where $\langle \sigma_{\rm eff}v\rangle \equiv \sqrt{\langle \sigma_{\rm SM\rightarrow \gamma'} v\rangle \langle \sigma_{\rm DM\rightarrow \gamma'} v\rangle}$). {This gives a $\Omega_{DM}\propto \log(\alpha'\epsilon)/{\alpha'}^2$ scaling, explaining} why this regime
leads to a line which is close to horizontal in the phase diagram of Fig.~\ref{fig::DogDiagram1}.

As said above, IIIa reannihilation regime is a new regime that is absent in the case of a massless {dark photon.}
In the latter case, the reannihilation regime which shows up is IIIb, { along which} the dark sector is populated through $\rm SM\rightarrow DM$ processes, see Ref.~\cite{Chu:2011be}.
In this case, the Boltzmann equations are the same as for the regime IIIa, trading the $\gamma_{\rm SM\leftrightarrow \gamma'}$ source term in Eq.~(\ref{eq:DRARP}) for a  $\gamma_{\rm SM\leftrightarrow DM}$ source term in Eq.~(\ref{eq:DRADM}).
For the massive dark photon case, we will see below that this regime occurs only when one includes the thermal corrections and when $m_{\gamma'}$ is much smaller than $m_{\rm DM}$, so that the phase diagram is approximately the same as for the massless case (i.e.~with the shape of a ``mesa'', displaying only Ia, IIIb, IVb, Va and Vb regimes). As explained  in Ref.~\cite{Chu:2011be}, the regime IIIb leads to an abundance $\Omega_{\rm DM}$ which scales as $ \log (\alpha' \kappa)/\alpha'^2$.

%%%%%%%%%%%%%%%%%%%%%%%%%%%%%%%%%%%
\subsection{Secluded freeze-out: regimes IVa and IVb}
\label{sec:secluded}

As we have seen, the reannihilation regimes occur when the slow source term producing hidden sector particles from SM particles is still active at the time when the temperature of the thermalized hidden sector, $T'$, becomes smaller than $m_{\rm DM}$.
If the source term  becomes inactive before $T'\sim m_{\rm DM}$, the system would not be in the reannihilation regime but, instead, the DM particles would undergo a simple ``secluded freeze-out''. By this, we mean standard,  text-books freeze-out, except that it takes place in a hidden sector, characterized by a temperature $T'$ which differs from that of the visible sector, $T$ (see Ref.~\cite{Chu:2011be} for details). 

In practice, though, secluded freeze-out does not occur in the instances depicted in Figs.~\ref{fig::DogDiagram1} and \ref{fig::DogDiagram2}.
This stems from the fact that, for the values of the masses considered for these figures, with in particular $m_{\gamma'}<m_{DM}$,
the $\rm SM\rightarrow \gamma'$ source term is  still active at $T'\sim m_{\rm DM}$. This is so because there is no IR mass scale which could cut-off this source term at $T'\gtrsim m_{\rm DM}$.  Nevertheless, there are  values of the masses for which the secluded freeze-out regime clearly occurs. 
This is in particular the case if $m_{\gamma'}< m_{\rm DM}<m_e$, as in this case the mass of the electron cuts the $\rm SM\rightarrow DM$ and $SM \rightarrow \gamma'$ source term at $T\sim m_e$, which can occur before $T'$ goes below $m_{\rm DM}$. In principle, one can distinguish two regimes here, depending on whether the dominant source term comes from the $\rm SM\rightarrow DM$ or $\rm SM \rightarrow \gamma'$ processes, corresponding to IVa and IVb regimes respectively. However, in practice, without taking into account thermal corrections, in this case only the regime IVa takes place  because the $\rm SM\rightarrow \gamma'$ source term naturally dominates
in the region of parameter space in the secluded regime. Thermal corrections change  this picture. As we shall see below,  the regime IVb is the one which applies in this case. In all cases,  the relic density is essentially determined by the value of $\alpha'$, leading to an approximately horizontal line in the phase diagram, see Fig.~\ref{fig::DogDiagram1} and Ref.~\cite{Chu:2011be}.

%%%%%%%%%%%%%%%%%%%%%%%%%%%%%%%%%%%%%%%%
\subsection{Freeze-out: regimes Va and Vb}

Finally, from the reannihilation or secluded freeze-out regimes, if we increase $\kappa$ (and consequently $\epsilon$), at some point all particles  form a single thermal bath, characterized by a single temperature $T$, so that DM undergoes a standard freeze-out.
This happens when, on top of the $\alpha'$ driven processes which were already in thermal equilibrium in the previous regimes, the $\kappa$ driven processes and/or the $\epsilon$ driven processes thermalize. Thus, such a transition into the freeze-out regime takes place when either $\kappa$ becomes larger than $\kappa_{\rm th}$, Eq.~(\ref{kappath}) or $\epsilon$ becomes larger than $\epsilon_{\rm th}$, Eq.~(\ref{epsilonth}).
Also, the freeze-out can be 
dominated either by the $\rm \gamma'\leftrightarrow DM$ annihilation process (leading to regime Va, depending only on the value of $\alpha'$, horizontal line in Fig.~\ref{fig:FIAndDFIPS}) or by the $\rm SM\leftrightarrow DM$ process for larger values of $\kappa$ (leading to regime Vb, which depends only on the value of $\kappa$, vertical line in Fig.~\ref{fig:FIAndDFIPS}).
The transition between regimes Va and Vb itself occurs when the $\rm SM\leftrightarrow DM$ rate becomes larger than the  $\rm \gamma'\leftrightarrow DM$ one.

%%%%%%%%%%%%%%%%%%%%%%%%%%%%%%%%%%%%%%%%
\section{Including thermal effects}
\label{sec:ThermalEffects}
%%%%%%%%%%%%%%%%%%%%%%%%%%%%%%%%%%%%%%%%

All  the discussion of DM production regimes above has been done without including any thermal effects
and is generic of what could also happen in other models of DM based on three sectors with three connectors. 
The dark photon model we considered is a bit special, as it is known for displaying specific thermal effects, which can be important in some cases. This has been studied at length in the context of dark photon production, in particular in stars \cite{Redondo:2008aa,Jaeckel:2008fi,Redondo:2008ec,An:2013yfc,Redondo:2013lna,Fradette:2014sza}. 
Based on these studies, we will discuss now these effects (focusing on dark photon production in the early universe) and on the change they imply for each DM production regimes unveiled in the previous section.

The most relevant issue concerns how to treat the limit $m_{\gamma'}\rightarrow~0$, starting from a massive dark photon. 
Indeed, in the massive case, one distinguishes the dark photon interaction and mass eigenstates bases (see Section \ref{sec:sectors} above), while for  $m_{\gamma'} \rightarrow 0$
the $A_\mu$ and $A'_\mu$ fields are degenerate and the distinction between the bases disappears. More concretely, this implies that,  in the massless limit,  dark photons do not couple to SM particles and so cannot be produced, see the discussion following Eq.~(\ref{eq:Lagrangian2}). How to take this effect properly into account is subtle but, in presence of a medium, as in the universe or inside of stars, this has been extensively studied in the literature and, eventually, clarified \cite{Redondo:2008aa,Jaeckel:2008fi,Redondo:2008ec,An:2013yfc,Redondo:2013lna,Fradette:2014sza}. As this problem is central for a proper incorporation of the thermal effects, we recap the key results of these works  in Appendix \ref{app:1}, while here we only summarize the salient points, working in the interaction basis. 

\begin{center}
\begin{figure}[h!]
\includegraphics[scale=.2]{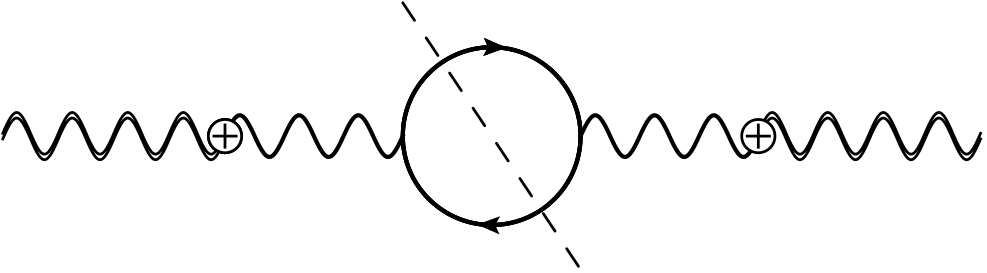}
\caption{The imaginary part of the dark photon propagator (double wiggly lines) in a medium includes both its decay rate and creation rates.}
\label{fig:fg1}
\end{figure}
\end{center}

In a thermal bath, a propagating dark photon can mix with an ordinary photon, which in turn interacts with the plasma of charged particles. This process is depicted in Fig.~\ref{fig:fg1}. In this figure, the double (single) wiggly lines depicts the dark photon  (resp. photons) propagators. The crossed circles represent their mixing $\propto \epsilon\, m_{\gamma'}^2$ and the blob is the photon polarization in the thermal bath, for instance made of relativistic $e^+e^-$ particles. The dashed straight line is a cut associated to taking the imaginary part of the photon polarization tensor. In vacuum, if non-zero, this cut would be related to the decay rate of the dark photon into, say $e^+e^-$ pairs, but in a plasma it includes also the  dark photon creation rate from coalescence processes, {\em i.e.} $e^+e^-\rightarrow \gamma'$ \cite{Redondo:2008aa,Weldon:1983jn}. By the same token, a cut in a two-loop diagram with photon exchange within the polarization tensor would lead to the dark photon creation rate from Compton scattering, say $\gamma \, e^\pm \rightarrow \gamma' \, e^\pm$, and pair annihilation, {\em i.e.} $e^+ \,e^- \rightarrow \gamma\, \gamma'$, etc. 

On top of this, in a medium one must distinguish the behavior of the transverse and of longitudinal  components of the photon polarization tensor.  The latter corresponds to genuine excitations of the medium, known as plasmons, see {\em e.g.} \cite{Bellac:2011kqa}. This, in turn, impacts the creation rate for production of transverse and longitudinal dark photons. As we discuss in Appendix \ref{app:1}, in most of the range of interest for dark matter particle production in the early universe, creation of dark photons is dominantly through transverse photons \cite{An:2013yfc,Redondo:2013lna}. In a thermal bath, the latter behave essentially as massive particles, with a thermal mass (here noted $\omega_T$) \cite{Braaten:1993jw}.
\begin{equation}
\operatorname{Re}\Pi_{\gamma,T} \equiv \omega^2_{T} \sim \sum_i e^2_i T^2,
\end{equation}
where $\Pi_{\gamma,T}$ is the self-energy of transverse photons and 
 $e_i$ is the electric charge of the relativistic particles present in the primordial plasma. Taking this into account, the creation of transverse dark photons proceeds through an effective mixing parameter $\epsilon \rightarrow \epsilon_{\rm eff}$, with 
\begin{equation} 
\label{eq:substitution}
\epsilon \rightarrow \epsilon_{\rm eff} = { \epsilon\, m_{\gamma'}^2 \over m_{\gamma'}^2 - \Pi_{\gamma,T}}.
\end{equation}
The numerator comes from the mixing mass term in Eq.~(\ref{eq:mixing_mass})  and the denominator from the transverse photon propagator with virtuality $k^2 = m_{\gamma'}^2$. In the resonance region $m_{\gamma'} \approx \omega_{\gamma,T}$, one must take into account the finite width of in-medium transverse photon modes, $\propto \operatorname{Im}\Pi_{\gamma,T} \ll \operatorname{Re}\Pi_{\gamma,T}$. Production at the resonance $m_{\gamma'} \sim \omega_{T}$ will play an important role in the sequel, but, essentially, the  substitution rule in Eq.~(\ref{eq:substitution})  implies that the production of dark photons is suppressed at high temperature/low dark photon mass, $m_{\gamma'} \ll \omega_{T}$. 
In particular, in the limit $m_{\gamma'}\rightarrow 0$, $\epsilon_{\rm eff}\rightarrow 0$, which means that there is no more $\rm SM\leftrightarrow \gamma'$ connection, so that the massless dark photon case is recovered smoothly ($i.e.$ if $m_{\gamma '}=0$, it can be rotated away and thus does not couple to SM EM currents).
This phenomenon  is familiar from the MSW effect in neutrino oscillations, in which oscillation into other neutrino flavors is suppressed by the frequent neutrino interactions in matter. In the opposite limit, however,  $\epsilon_{\rm eff} \rightarrow \epsilon$, and the production of dark photon is only suppressed by the bare mixing $\epsilon$ parameter, as in vacuum.\footnote{This is all there is to if the dark photon gets its mass through the St\"uckelberg mechanism. The case of spontaneous symmetry breaking is discussed in Appendix \ref{sec:BEH}.} Notice also, the substitution of Eq.~(\ref{eq:substitution}) is  for production of (transverse) dark photons on mass-shell and {\em not} in processes with exchange of virtual photons/dark photons, which are essentially insensitive to the dark photon mass. The rational behind Eq.~(\ref{eq:substitution}) is discussed in Appendix \ref{app:1}.

{The main practical consequences of the above discussion for the production of massive dark photons and DM in the early universe is that the dark photon production is suppressed at high temperature (compared to its mass) and subsequently is resonantly enhanced once $\omega_T\approx m_{\gamma '}$. The latter occurs at a temperature approximately equal to $T_{\rm res}\simeq \left(\text{a few}\right)m_{\gamma '}$. This implies that if $m_{\rm DM}$ is larger than $m_{\gamma'}$ by more than one to two orders of magnitude, i.e.~ if $m_{\rm DM}\gg T_{\rm res}$, there is basically no significant production of dark photons from the SM at times that are relevant for DM production,
so one recovers a mesa-shaped phase diagram similar to the massless case. Thus regimes Ib, II, IIIa and IVa are irrelevant and only regimes Ia, IIIb, IVb, Va and Vb remain, as in the massless case. 

\begin{figure}[h!]
\subfloat[]{\includegraphics[height=0.8\linewidth]{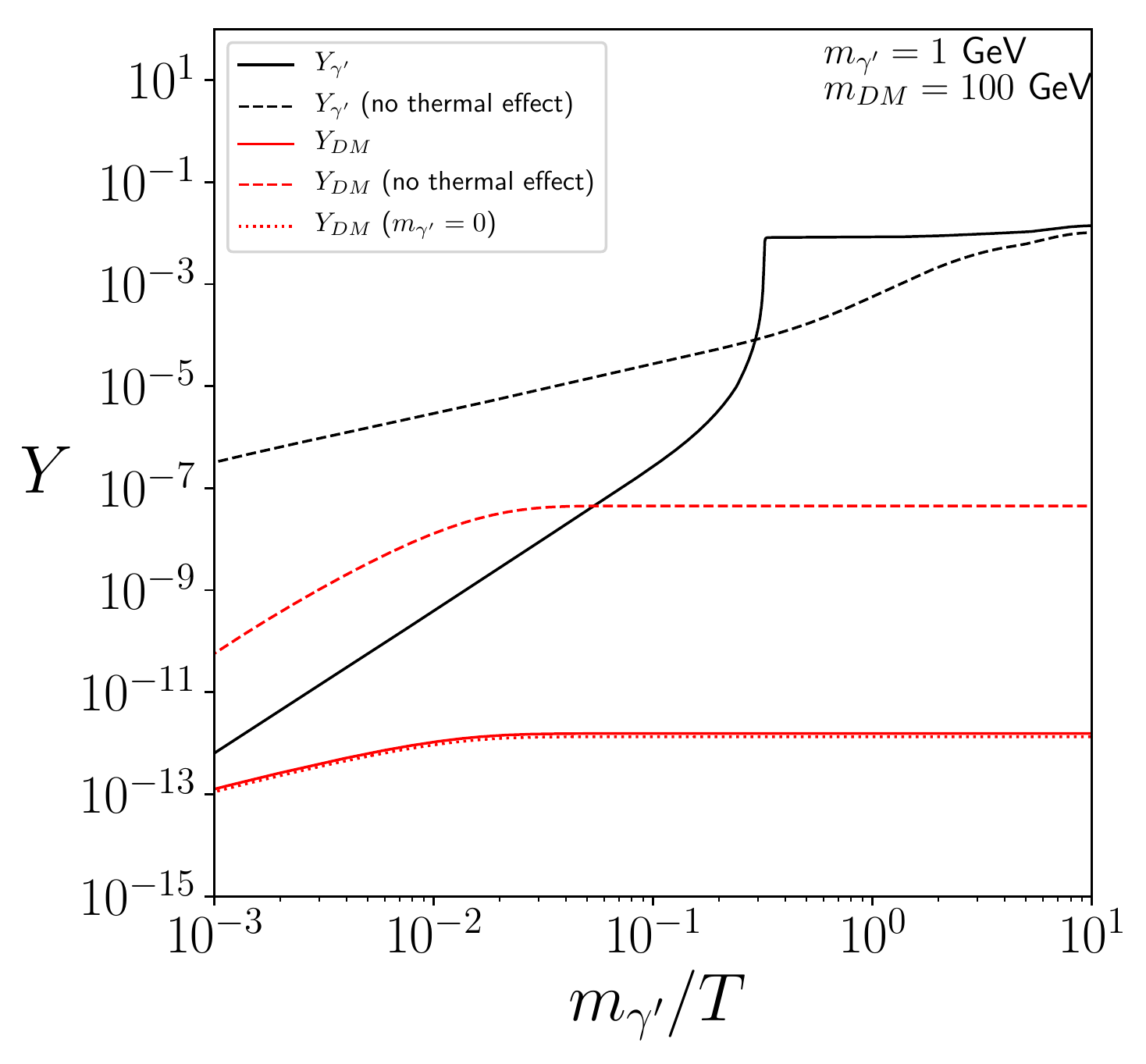}}\\
\subfloat[]{\includegraphics[height=0.8\linewidth]{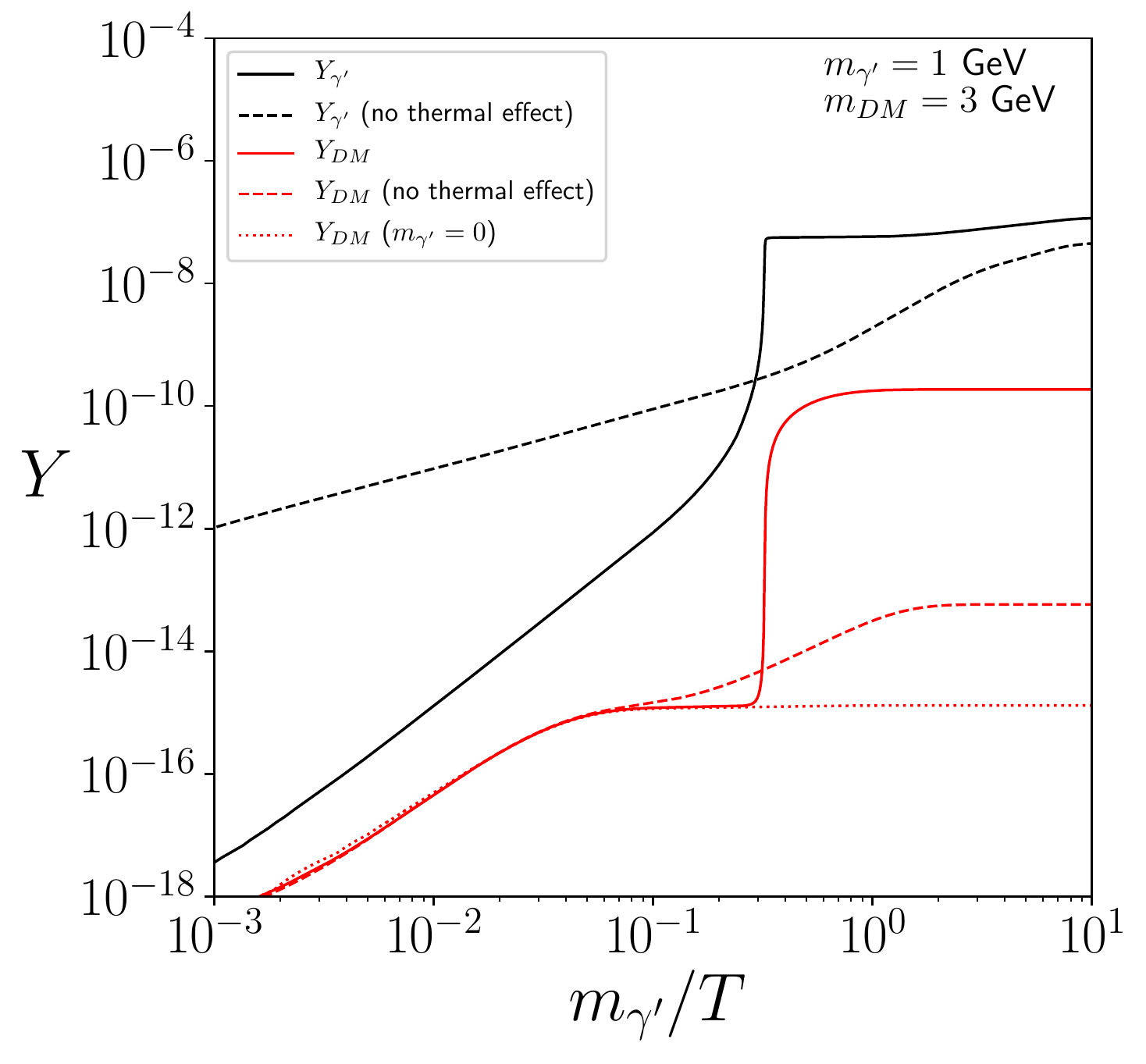}}\\
\caption{Comparison of the evolution of the dark photon (black lines) and DM abundances (red lines), with (solid) and without (dashed) thermal effects on dark photon production. The dotted lines give the abundance of dark photon in the massless limit. Panel (a) and (b) differ by the choice of dark matter and dark photon masses. {Both panels consider values of the couplings which lead to the observed relic density when the thermal corrections are taken into account (solid red lines): $\kappa=2\times 10^{-11}$ and $\epsilon=5.6 \times 10^{-9}$ for panel (a) and $\kappa_{\rm II}=3.6\times 10^{-14}$ and $\epsilon = 10^{-11}$ for panel (b).}}
\label{fig::Yields}
\end{figure}

Fig.~\ref{fig::Yields} gives the evolution of the dark photon and DM yields, $Y_{\gamma'}$ and $Y_{\rm DM}$ as a function of $m_{\gamma'}/T$ and for two characteristic sets of parameters. 
The solid (dotted) lines shows the evolution including (resp. not including) the thermal corrections to dark photon production (see Appendix \ref{app:1} for details). The dotted lines show the evolution of $Y_{\rm DM}$ taking $m_{\gamma'}=0$ with other parameters unchanged.

In the first panel (a) we have taken $m_{\gamma '}=1$ GeV, $m_{\rm DM}=100$ GeV and couplings that lead to the observed relic density when we include the thermal effects ($\kappa=2\times 10^{-11}$ and $\epsilon=5.6 \times 10^{-9}$).}
Given that DM is two orders of magnitude heavier than the dark photon, this panel shows  that, as expected, the production of dark photons is suppressed at high temperatures, and in particular at temperatures that are relevant for DM production, $T\gtrsim m_{\rm DM}$.
As a result when the dark photon yield becomes important, around $T_{\rm res}\simeq m_{\gamma '}/\left(\text{a few}\right)$, the production rate of DM from dark photon pair annihilation is already too much Boltzmann suppressed to play any sizeable role in the final amount of DM (red solid line). This means that DM is mostly produced directly through $\kappa$ driven $\rm SM\rightarrow DM$ processes and that the relic density  is very close to the one obtained considering a massless dark photon (red dotted line). Thus the value of $\kappa$  is close to the value one needs in the massless case,
$\kappa _{\rm II}\simeq \kappa_{\rm Ia}=2\times 10^{-11}$. 
If, instead, one considers the dark photon as massive but neglect the thermal corrections, there is no suppression of the dark photon production at high temperatures and DM is sizeably produced through dark photons. As a result, one obtains a larger DM relic density (red dashed line). Therefore, without thermal corrections, one would need  a smaller value of $\kappa$. Indeed, for these masses, regime II applies so that the overall scaling of the dark photon production of DM is $\propto \epsilon^4\alpha'^2\propto \kappa^4 $, in which case one would need $\kappa _{\rm II}=2\times 10^{-12}$.

The second panel (b) illustrates the opposite situation, where the dark photon and DM  masses are closer, $m_{\gamma '}=1$ GeV, $m_{\rm DM}=3$~GeV. Again this is shown for values of the couplings that lead to the observed relic density taking into account thermal corrections, $\kappa_{\rm II}=3.6\times 10^{-14}$ and $\epsilon = 10^{-11}$.
With thermal effects included, the resonance appears at a temperature close to $m_{\gamma '}\sim m_{\rm DM}$. As the production rate of DM from dark photon is not Boltzmann suppressed when the dark photon production is resonantly enhanced, the DM abundance strongly increases, following the behavior of the dark photon abundance. Consequently, the thermal corrections  strongly enhance the DM production from dark photon pair annihilation, and this both with respect to the massive case without thermal correction and with respect to the massless case.
This implies that the {value of $\kappa$ needed} with thermal corrections is smaller than the one required in the massless case, $\kappa_{\rm Ia} \approx 10^{-11}$, and is also smaller than the one { needed} in the massive case without thermal corrections, $\kappa _{\rm II}=3\times 10^{-13}$. In the latter case, the production is not suppressed at high temperatures but is not boosted afterwards by any resonance in the production. See Appendix~\ref{app:1} for a more detailed study of the effect of the resonance on the DM production. 
Many of the features just explained can also be read off Fig.~\ref{fig:FIAndDFIPSThermal} which, as Fig.~\ref{fig:FIAndDFIPS}, gives the value of $\kappa$ as a function of $m_{\rm DM}$ for $m_{\gamma'}=1$~GeV and $m_{\gamma'}=10$~GeV, but including the thermal corrections.

One related general consequence of thermal effects is that they affect the thermalization of the dark photons with the SM thermal bath. As a result, the simple criterion $\Gamma/H>1$ of Eq.~(\ref{eq:DPThermSM}) has to be corrected, as the production can be resonantly enhanced around $T\sim T_{\rm res}$ (see Eq.~(\ref{eq:substitution}) and Appendix \ref{app:1}).
Thus, taking into account thermal effects, to determine whether thermalization occurs (i.e.~if $n_{\gamma'}$ reaches $\sim n_{\gamma'}^{\rm eq}(z)$) one needs to solve explicitly the Boltzmann equations. In this way, we find that the thermalization condition becomes
\begin{align}
\epsilon\gtrsim \epsilon_{\rm th}\equiv 6\times 10^{-9}\left(\frac{m_{\rm DM}}{\text{GeV}}\right)^{1/2}.
\label{epsilonthTE}
\end{align}
Comparing with Eq.~(\ref{eq:DPThermSM}), we see that, as expected, a smaller {value of $\epsilon$} is needed for the dark photons to thermalize with the SM sector.

\begin{center}
\begin{figure}[h!]
\includegraphics[scale=0.6]{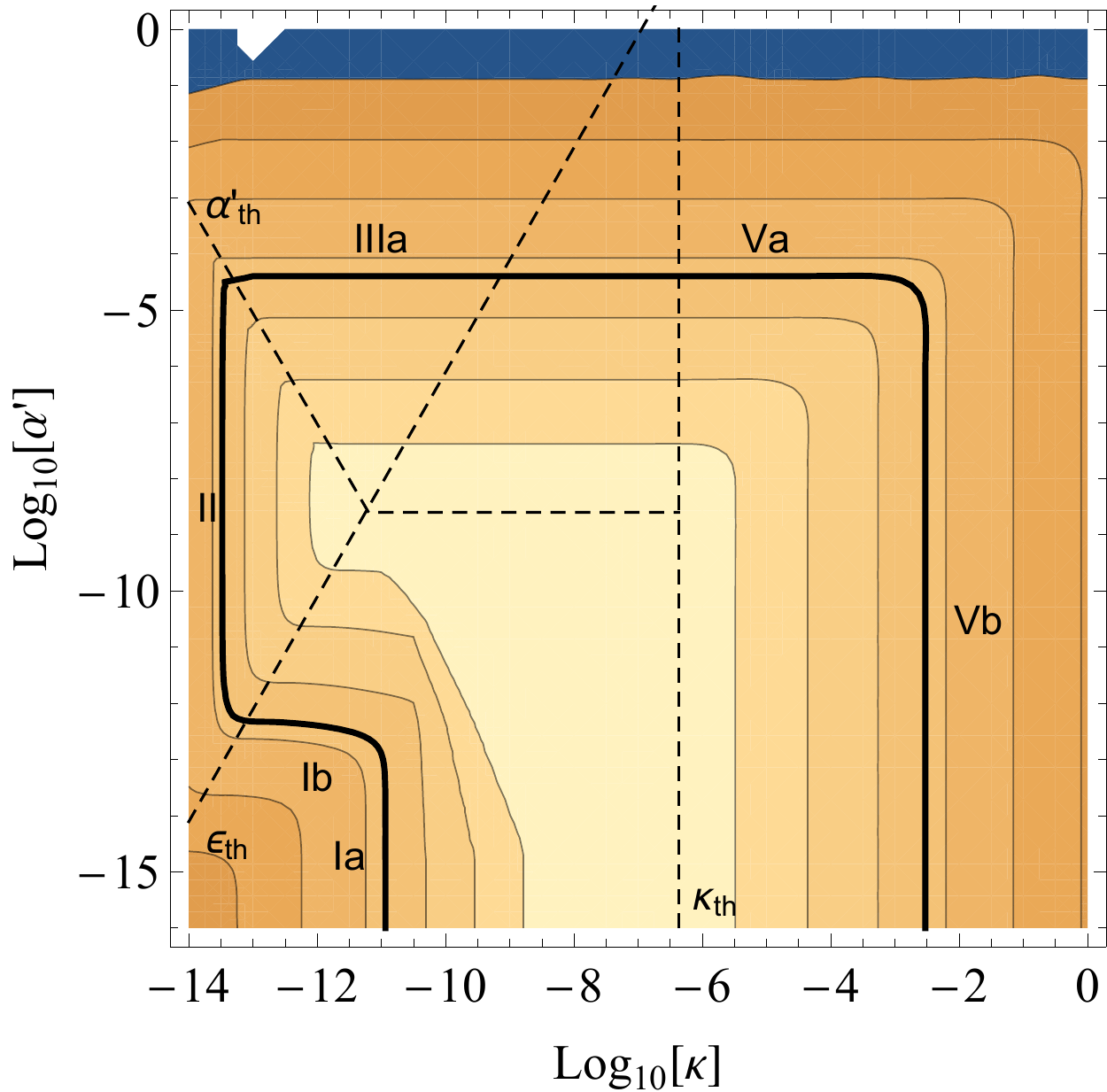}
\caption{Same as Fig.~\ref{fig::DogDiagram1},  $m_{\rm DM} = 3 $ GeV and $m_{\gamma'} = 1$ GeV, but taking into account thermal effects on dark photon production.}\label{fig::MESA_3_1} 
\end{figure}
\end{center}

\begin{center}
\begin{figure}[h!]
\includegraphics[scale=0.6]{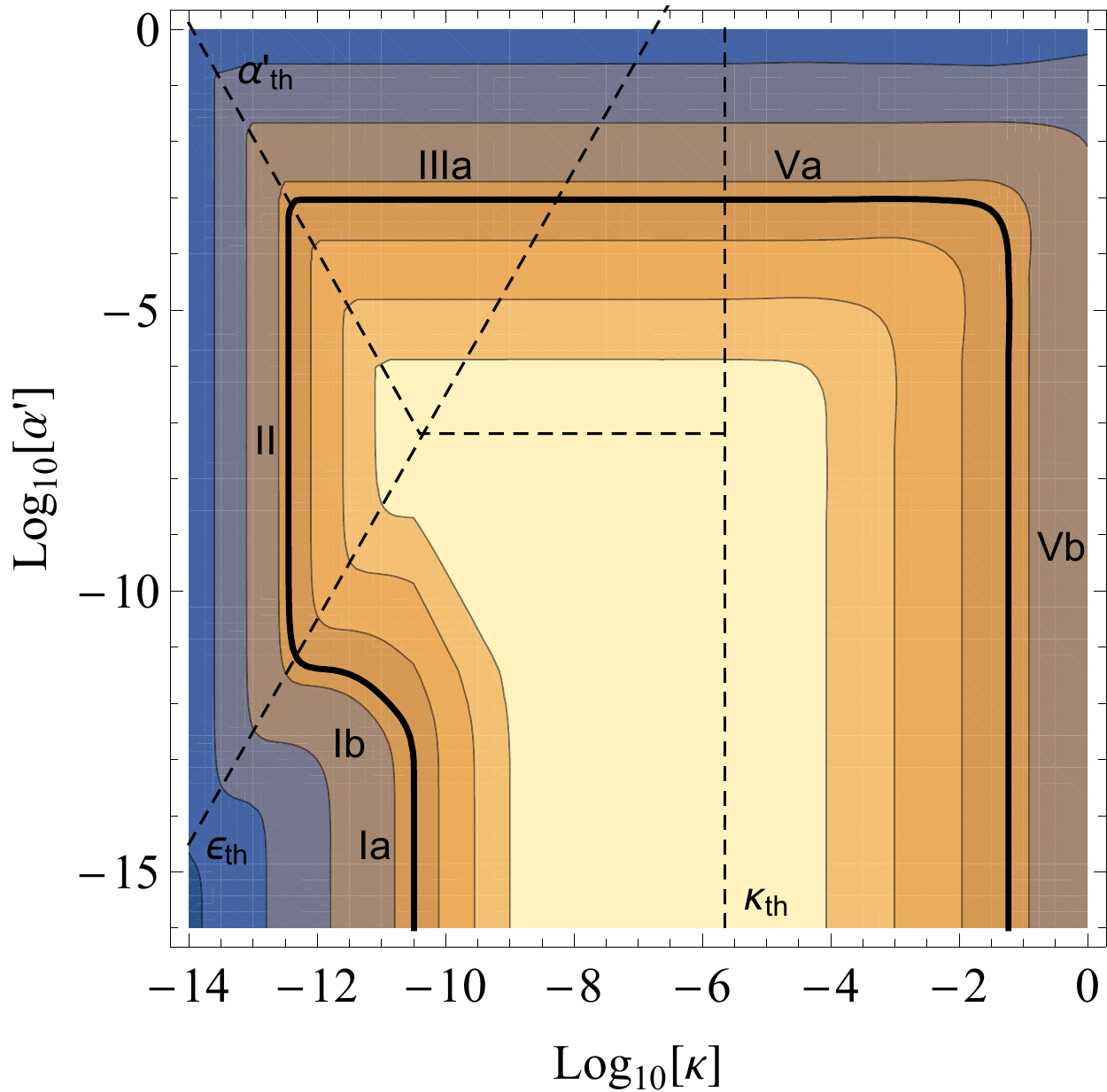}
\caption{Same as Fig.~\ref{fig::DogDiagram2},  $m_{\rm DM} = 100 $ GeV and $m_{\gamma'} = 10$ GeV, but with thermal effects included.}\label{fig::MESA_100_10}
\end{figure}
\end{center}

\begin{figure}
\includegraphics[width=8cm]{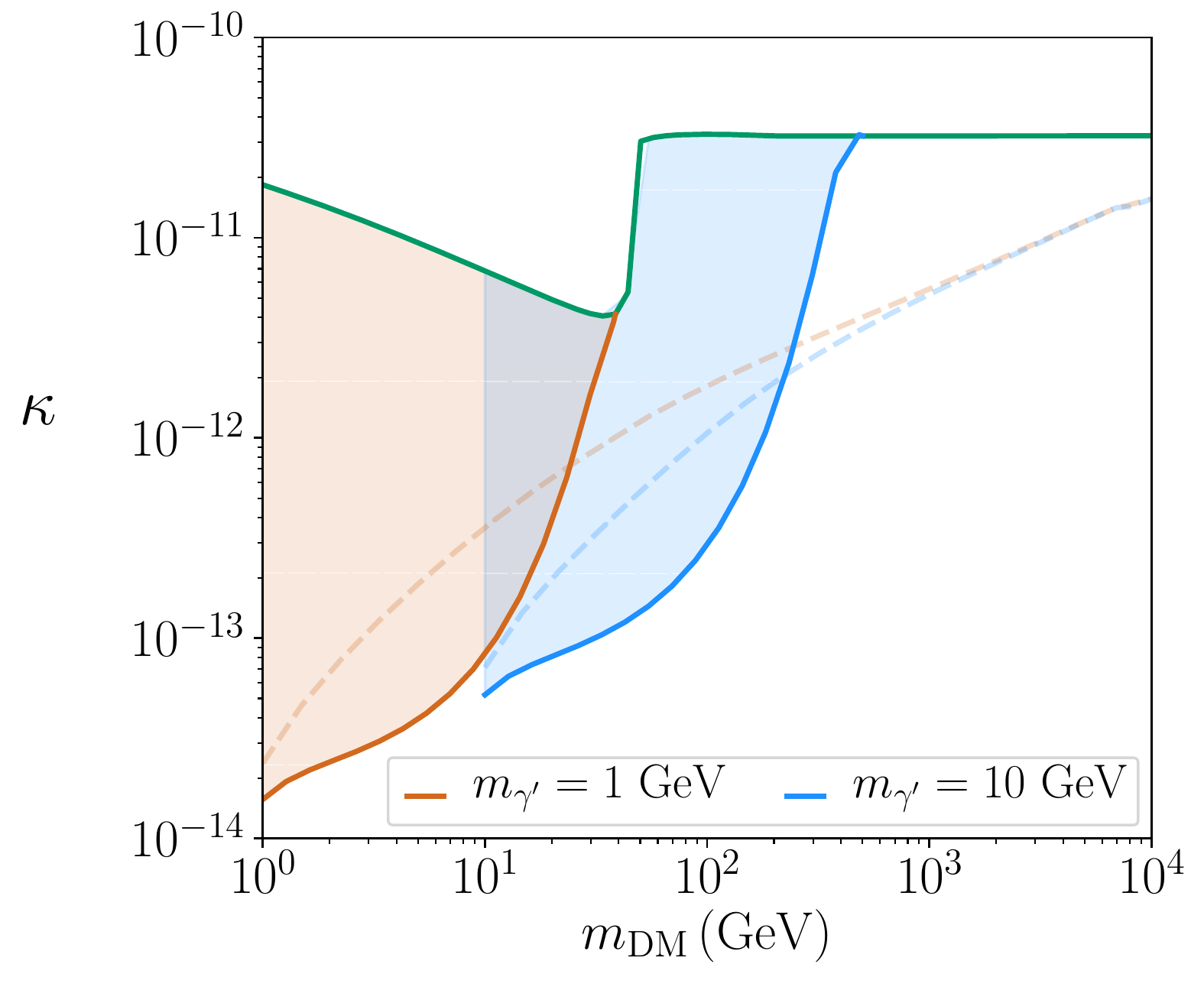}
\caption{Dark matter candidates as shown in Fig.~\ref{fig:FIAndDFIPS} but with thermal corrections to dark photon production taken into account. The additional dashed lines are given for the sake of comparison. They correspond to the curves shown in Fig.~\ref{fig:FIAndDFIPS}, thus without thermal effects.}
\label{fig:FIAndDFIPSThermal}
\end{figure}

Fig.~\ref{fig::MESA_3_1} gives the phase diagram obtained including the thermal effects, for the same value of $m_{\rm DM}$ { and $m_{\gamma'}$ than in Fig.~\ref{fig::DogDiagram1}, i.e.~$m_{\gamma '}=1$ GeV, $m_{\rm DM}=3$~GeV.}
One observes that, as expected from Fig.~\ref{fig::Yields}, the "mooring bollard" shape of the phase diagram is accentuated by the thermal corrections, as the gap between the values of $\kappa$ in the Ia and II regimes is larger.
Similarly, Fig.~\ref{fig::MESA_100_10} gives the phase diagram including the thermal effects for the same values of the masses as in Fig.~\ref{fig::DogDiagram2}.

To explain the various changes in these phase diagrams, we now revisit each regime one by one.

\noindent\underline{{Freeze-in regimes}}: In the Ia regime, the thermal effects are small, simply because  the $\rm SM\rightarrow \gamma'$ processes are irrelevant in this regime. In regime Ib instead, the approximation we used of taking $n^{\rm eq}_{\gamma'}(z)$ in the Boltzmann equation of Eq.~(\ref{BoltzEqIb}) does not give always the correct result. Instead, one has to take the actual value of $n_{\gamma'}$, as determined from Eq.~(\ref{BoltzEqII2}). If $m_{\rm DM}\gg m_{\gamma '}$, Eq.~(\ref{BoltzEqII2}) will give $n_{\gamma '}\approx 0$ and the Ib regime will disappear. Instead,  if $m_{\rm DM}\gtrsim m_{\gamma '} $, $n_{\gamma '}$ will be enhanced by the resonance and regime Ib will be relevant for a larger range of couplings. The latter case is manifest in both phase diagrams with thermal effects. If the resonance is at $T<m_{\rm DM}/3$ ($i.e.$ intermediate DM masses), the $\rm \gamma '\gamma '\rightarrow DM DM $ freeze-in process is Boltzmann suppressed and so is not as efficient as assumed when deriving the $\alpha'_{\rm Ia}$ coupling. Thus, the turn towards smaller value of $\kappa$ can happen for larger values of $\alpha'$, see Fig.~\ref{fig::MESA_100_10} for such an example.

\noindent \underline{{Sequential freeze-in}}: for $m_{\gamma'}$ not much smaller than $m_{\rm DM}$, as in Figs.~\ref{fig::MESA_3_1}  and \ref{fig::MESA_100_10}, the thermal effects lead to smaller values of $\kappa$ than without thermal corrections (see the discussion on the second panel of Fig.~\ref{fig::Yields}
which gives the observed relic density along this regime).
With thermal corrections, this regime leads to a vertical line in the phase diagram (as in the case without thermal corrections) because the $Y_{\gamma '}\propto \epsilon^2 $ and $\gamma_{\rm \gamma '\leftrightarrow DM}\propto \alpha'^2$ scalings
remain unchanged.
{For $m_{\gamma'}\ll m_{\rm DM}$ this regime and the intermediate regime Ib merge with the standard Ia freeze-in regime, as shown in Fig.~\ref{fig:FIAndDFIPSThermal} for two choices of dark photon masses (solid lines). Compared to Fig.~\ref{fig:FIAndDFIPS} (reproduced as dashed curves), the noticeable features are that resonant production of dark photon leads to even smaller values of $\kappa$ (see the discussion revolving around Fig.~\ref{fig::Yields}) and that the regimes II (orange and blue curves) and Ia (green curve) merge for relatively smaller {values of $m_{DM}/m_{\gamma'}$, for} reasons explained above.}

\noindent\underline{{Reannihilation regimes}}: for $m_{\gamma'}$ not much smaller than $m_{\rm DM}$, as for Figs.~\ref{fig::MESA_3_1}  and \ref{fig::MESA_100_10}, we have seen above that the production of DM is boosted from the fact that the production of $\gamma'$ is resonantly boosted at a temperature which is still relevant for DM production. Thus the IIIa  production regime, i.e. DM through $\gamma'$, dominate even more over $\rm SM\rightarrow DM$ production than in the case without thermal corrections (where it was already dominant, see above). Instead, for $m_{\gamma'}\ll m_{\rm DM}$ regime IIIb, where $\rm SM\rightarrow DM$ processes dominates, applies as in the massless case.

\noindent\underline{{Secluded freeze-out}}: for $m_\gamma'< m_{\rm DM}<m_e$, the IVa regime, which applies in absence of thermal effects, see above, does not apply anymore as in this case the production of dark photons is highly suppressed.
This is due to the fact that, at $T \gg m_{\gamma'}$, the $\gamma'$ productions is suppressed by thermal corrections and that  for $T\sim m_{\gamma'}<m_e$ there is no sizeable resonant production as  the production is anyway Boltzmann suppressed.
Regime IVb nevertheless can apply for $m_{\gamma'}\ll m_{\rm DM}$, when production from the SM model is dominated by $Z$-decays, as in the massless case, see Ref.~\cite{Chu:2011be}. 
There is no IIIa to IVb transition (but only IIIa to Va transition) because, as $\kappa$ increases, so does $\epsilon$ and thus the production of dark photons. 

\noindent\underline{{Freeze-out regimes}}: since  all particles are thermalized in a single thermal bath, the effects of the above thermal corrections on the dynamics of these regimes are negligible. 

%%%%%%%%%%%%%%%%%%%%%%%%%%%
\section{Comments on constraints}
\label{sec:constraints}

As the goal of our work is primarily the classification of the ways of producing DM, we only briefly comment on possible constraints, with a focus on the harder to test, { small $\kappa$} regimes. Indeed,  as is well-known, much of the parameter space corresponding to the freeze-out and reannihilation regimes of  model (\ref{eq:model}) are already constrained or excluded, including by the direct detection searches, see e.g. \cite{Chu:2011be,Hambye:2018dpi}. For the parameter ranges relevant for the freeze-in regimes, no  consequent productions at colliders and indirect detection experiments of hidden sector particle is expected in general. The rates of processes relevant for those searches, $i.e.$ $\rm SM\rightarrow \gamma'$ or $\rm SM\rightarrow \gamma' \rightarrow DM$ processes are typically strongly suppressed by the tiny parameters. Even so, if the mediator is light enough (meaning  $m_{\gamma '}<20\text{ MeV} $ in the case of direct detection searches of DM in the $m_{\rm DM} \gtrsim$ GeV and above range),  Rutherford scattering of DM on nuclei may be strongly enhanced, so that for some values of the masses direct detection experiments are already testing the freeze-in regime \cite{Hambye:2018dpi,Chu:2011be}. For lighter dark matter candidates, constraints on the parameter space of the freeze-in regime of millicharged dark matter will be possible in the future, exploiting scattering of DM off electrons, provided the dark photon is light enough,  $m_{\gamma'} \lesssim$ keV  \cite{Essig:2011nj}. Sequential freeze-in, as well as freeze-in regime Ib, in which DM is produced by the mediator, here the dark photons, extend the range of viable DM candidates toward even  more feeble DM to SM couplings. In the case of the dark photon, the existence of these regimes is correlated to the mass of the dark photons, and in particular to the thermal effects discussed in Section \ref{sec:ThermalEffects}. Specifically, for dark photons much lighter than DM, $m_{\gamma'} \ll m_{\rm DM}$, the production of DM is effectively as in the massless dark photon regime, and so along the standard Ia freeze-in regime. This does not precludes the possibility that the new freeze-in regimes could also be tested by direct detection, including for other models than (\ref{eq:model})  \cite{wip}. 

Light dark photons and light millicharged particles are also constrained by stellar as well as CMB (cosmic microwave background) and BBN (big bang nucleosynthesis) measurements. In particular, the impact of massive, feebly coupled dark photon produced by freeze-in on cosmological observables has been studied in details \cite{Berger:2016vxi} (see also \cite{Fradette:2014sza}). BBN can constrain dark photons that have an hadronic decay channel open, $m_{\gamma '}>2\times m_{\pi}\sim 300\text{ MeV} $ while CMB measurements are relevant for lighter dark photons $2m_{e} \lesssim m_{\gamma '} \lesssim 100\text{ MeV} $, through their energy injection from dark photon decay into $e^+/e^-$. For instance, if $m_{\gamma '}=1\text{ GeV}$, as in the phase diagram of Fig.~\ref{fig::MESA_3_1},  then the range  $10^{-12}\lesssim \epsilon \lesssim 10^{-10}$, which should show as an oblique band in the phase diagram of Fig.~\ref{fig::DogDiagram1}, is excluded by BBN, as can be seen from Fig.~5 of Ref.~\cite{Berger:2016vxi}. At the same time, for $m_{\rm DM} = 3$ GeV, the sequential freeze-in runs from  $10^{-12} \lesssim \epsilon_{\rm II} \lesssim 3\times 10^{-7}$, roughly, so that several candidates whose abundance are set by sequential freeze-in are not excluded by BBN measurements (or, for that matter, by any other kind of currently known constraints).  
For $m_{\rm \gamma'} = 10$ GeV, the range that is constrained by BBN corresponds to $5 \times 10^{-11} \lesssim \epsilon \lesssim 5\times 10^{-12}$ and $10^{-14} \lesssim \epsilon \lesssim 10^{-13}$, while for and $m_{\rm DM} = 100$ GeV,  $3\times 10^{-12} \lesssim \epsilon_{\rm II} \lesssim 10^{-7}$, again leaving room for viable candidates in the sequential freeze-in regime.

%%%%%%%%%%%%%%%%%%%%%%%%%%%%%%%%%%%%%%%%%

%%%%%%%%%%%%%%%%%%%%%%%%%%%%%%%%%%%%%%%%%

\section{Generalization to other models}
\label{sec::gen}

The model we considered above has a particular structure where the three connections are determined by only two independent parameters. 
One could therefore wonder if the overall DM production regime picture we have obtained above is deeply connected to this particular structure or is more general. In this section we would like to stress that this picture applies to other models and that in particular the sequential freeze-in regime could lead to the observed relic density in many setups. In some cases it can even be the only possible way to produce DM.

To illustrate this discussion, let us take another example, where DM is a fermionic particle $\chi$, coupled to a scalar particle $\phi$,
\begin{equation}
{\cal L }\owns -(Y_\chi \bar{\chi}\chi \phi + h.c.)-\lambda_\phi \phi^\dagger \phi H^\dagger H. \label{Lagrscalarcase}
\end{equation}  
From such a structure, if the $\phi$ scalar field does not acquire any vev, clearly one cannot create any DM particle directly from the SM (at lowest {order in $Y_\chi$ and $\lambda_\phi$).} The {way to produce DM from the SM is then} to create first $\phi$ particles and from there DM particles. Thus the three new regimes we presented above, Ib, II, IIIa and IVa are the generic possible ones, with $Y_\chi$ and $\lambda_{\phi}$
playing the role of $e'$ and $\epsilon$ in the kinetic mixing model above, respectively. In particular if none of the interaction thermalize the sequential freeze-in is the only possible one.
If instead the $\phi$ scalar field has a vev, it mixes with the Higgs boson and one gets a structure very similar to the one of the kinetic mixing model above, with 2 parameters inducing the 3 connections between the SM, $\phi$ and $\psi$ fluids.
DM can be pair produced from SM particles through Higgs and $\phi$ exchange, from a combination of the $Y_\chi$ and $\lambda_\phi$ interactions (similar to the $\kappa$ combination above), $\phi$ can be produced from the SM through $\lambda_\phi$ (similar to $\epsilon$ above) and DM particles can be produced by $\phi$ particles through the $Y_\chi$ interaction (similar to the $e'$ coupling above).\footnote {Note that, unless it is forbidden in various ways, we could also have a $\phi H^\dagger H$ interactions which implies a $\phi-h$ mixing too. In presence of such a term the various connections also depend on this interaction.}   Thus one expects a phase diagram similar to the one we obtained above.    

The structure of the phase diagram of Fig.~\ref{fig::DogDiagram1} is deeply connected to the fact that the sizes of the SM to mediator and mediator to DM interactions implies a lower bound on the size
of the processes which connect directly SM and DM particles. For the kinetic mixing case this is quantified by the $\kappa=\epsilon\sqrt{\alpha'/\alpha}$ relation.
For the scalar mediation example of Eq.~(\ref{Lagrscalarcase}) a similar phenomena occurs. If instead one would have three totally independent connections between the three sectors,
still the various regimes encountered above would be possible, in particular the sequential freeze-in, but the overall picture would become more complicated, as the phase diagram becomes a 3-d diagram, rather than a 2-d one, as in Fig.~\ref{fig::DogDiagram1}.\footnote{A simple possibility of this kind would be to have a scalar mediator $\phi$ and a scalar DM particles $\phi_{\rm DM}$ with 3 quartic terms, ${\cal L}\owns -\lambda_{H-\phi} \phi^\dagger \phi H^\dagger H -\lambda_{\rm{DM}-H} \phi_{\rm DM}^\dagger \phi_{\rm DM} H^\dagger H -\lambda_{\rm DM-\phi} \phi_{\rm DM}^\dagger \phi_{\rm DM} \phi^\dagger \phi$. If both $\phi$ and $\phi_{\rm DM}$ have no vev, the 3 connections remain fully independent at tree level { and lowest order in the couplings}. Thus one has still a minimum size for one connection as a function of the other 2, but loop suppressed, which somewhat changes the overall picture.} In this case one can produce dark matter in the 9 ways described above.
Thus at least 9 regimes are possible, along the 5 different dynamical ways of producing DM from the SM, already mentioned above: freeze-in, sequential freeze-in, reannihilation, secluded freeze-out and freeze-out.

\section{Conclusions}
\label{sec:conclusions}

In this work, we have analysed the problem of DM production in the framework of a very popular model in which DM is charged under a hidden sector $U(1)'$ and interacts with SM particles
through kinetic mixing.
It is impressive that such a simple framework can lead to a structure of DM production regimes as rich as the one we have uncovered. Indeed, we have found that DM production out of the SM thermal bath through the kinetic mixing portal involves no less than 5 different dynamical ways, and this along 9 distinct regimes! 
This result, which we obtained considering the dark photon to be massive, generalizes the massless dark photon case for which 4 dynamical ways, along 5 regimes, were operative \cite{Chu:2011be}. The 4 new regimes, including the one based on a new DM production mechanism which we dub ``sequential freeze-in'' (regime II in our classification of possible regimes), all involve prior production of dark photons by SM particles. 
While we focused on the kinetic mixing portal model, the structure we have found is actually characteristic of DM setups in which, on top of the SM particles and DM particles, there are other particles, akin to the dark photons, which can couple both to the SM and DM particles.
More precisely, it is characteristic of models in which these 3 fluids  communicate along 3 connections, which are due to 2 different interactions, here $\alpha'$ and $\epsilon$.
Such a structure thus applies to many possible DM models.
Despite of the relative complexity of the DM production structure, a simple picture emerges in the phase diagram, as depicted in the plane $\alpha'$ vs $\kappa = \epsilon \sqrt{\alpha'/\alpha}$. Indeed, the different production regimes follow a universal pattern composed of vertical and (almost) horizontal lines in the phase diagrams, arranged in a shape suggesting a "mooring bollard". 

 In practice, the 4 new DM production regimes  turn out to be operative when the  mediator, here the dark photon, is lighter than the DM particles by a factor $m_{\gamma'}/m_{\rm DM} \gtrsim 10^{-2}$. Below this bound, the mass of the hidden photon is effectively negligible and one recovers the regimes of DM production of the massless case studied in details in Ref.~\cite{Chu:2011be}. In particular, one recovers in this limit the characteristic  "mesa" shape in the phase diagrams.  

The new ``sequential freeze-in'' dynamical production mechanism (regime II) involves the slow, out-of-equilibrium production of dark photons, followed by a slow out-of-equilibrium production of DM from these dark photons.
We find interesting that such a chain of freeze-in production of particles is operative and can lead to the observed relic density even in a model as simple as the kinetic portal.
The other new regimes are DM freeze-in production from a thermalized population of dark photon (Ib), reannihilation (IIIa) and secluded freeze-out (IVa)  through freeze-in production of dark photons.
For these new regimes, the thermal effects on dark photon production turn out to be important. On the one hand, they suppress the production of dark photons, and thus of DM, at temperature above the dark photon mass. On the other hand, they imply subsequently a resonant production of dark photon, which can strongly boost the DM production. 

\begin{acknowledgments}
This work  is supported by the F.R.S.– FNRS under the Excellence of Science (EoS) project No. 30820817 – be.h “The H boson gateway to physics beyond the Standard Model”, by the FRIA , the ”Probing dark matter with neutrinos” ULB-ARC convention and the IISN convention 4.4503.15. T.H. and L.V. thank the Erwin Schr\"odinger International Institute for hospitality while this work was completed.
\end{acknowledgments}

\appendix

\section{Thermal effects on dark photon production}
\label{app:1}

 The production rate of dark photons in a medium has been studied extensively, in particular in the sequence of articles  \cite{Redondo:2008aa,Jaeckel:2008fi,Redondo:2008ec,An:2013yfc,Redondo:2013lna,Fradette:2014sza}. For comprehensiveness, we recap here the main steps of their arguments.  We work in the interaction basis, treating mixing as a perturbation. Through mixing with SM photons, dark photons have a self-energy that inherits thermal features from those of photons, see Fig.~\ref{fig:fg1},
 \begin{equation}
 \Pi_{\gamma'} = m^2_{\gamma'} + {\epsilon^2 m_{\gamma'}^4\over (K^2 - \Pi_{\gamma})},
 \end{equation} 
 with $K^2 = \omega^2 - k^2$, where $K$ is the  momentum of the virtual dark photon. 
 In particular, its imaginary part is 
 \begin{equation}
\operatorname{Im} \Pi_{\gamma'}= {\epsilon^2 m_{\gamma'}^4 \operatorname{Im}\Pi_\gamma \over (m_{\gamma'}^2- \operatorname{Re}\Pi_{\gamma})^2 +  \operatorname{Im}\Pi_{\gamma}^2},
 \end{equation} 
 where we set $K^2 = m_{\gamma'}^2$ corresponding to mixing of a virtual photon into a dark photon on mass-shell. In vacuum, $ \operatorname{Im}\Pi_\gamma$ is related to possible photon decay channels, but in a thermal bath, $ \operatorname{Im}\Pi_\gamma$ has terms corresponding to both the emission and absorption rates of  photons from the medium\footnote{More precisely, $\Gamma$ is to be interpreted as the rate that determines the approach to thermal equilibrium, 
 $$
f(\omega,t) - f_{\rm eq} \propto \exp(- \Gamma t),
$$
where $f$ is the distribution function of (here) dark photons. The strange minus sign between the rates of emission and of absorption can be traced to Bose-Einstein statistics; for fermions $\Gamma =  \Gamma_{\rm em} + \Gamma_{\rm abs}$ \cite{Weldon:1983jn}.}
 \begin{equation}
\operatorname{Im}\Pi_\gamma = - \omega \Gamma_\gamma = - \omega (\Gamma_{\rm \gamma, \,em} - \Gamma_{\rm \gamma, \,abs}).
 \end{equation}
By detailed balance\footnote{More generally, to the fact that the amplitude squares for the emission and absorption process are equal, due to unitarity \cite{Weinberg:1979bt}.}, the emission rate of a photon of energy $\omega$ is Boltzmann suppressed compared to the corresponding absorption rate, 
\begin{equation}
\Gamma_{\rm em} = \exp{(- \omega/T)} \;\Gamma_{\rm abs},
\end{equation}
so that
\begin{equation}
\label{eq:emmissionrate}
\Gamma_{\rm \gamma',\, em} =   {\epsilon^2 m_{\gamma'}^4\Gamma_{\rm \gamma,\, em}  \over (m_{\gamma'}^2- \operatorname{Re}\Pi_{\gamma})^2 +  \omega^2(e^{\omega/T}-1)^2 \Gamma_{\rm \gamma,\, em}^2}.
\end{equation}
Now, at finite temperature, on top of genuine transverse photons, the medium can also sustain the propagation of longitudinal modes or plasmon waves. These correspond to  oscillations of charged particles that composed the thermal bath. Thus, in the expression of the rate of production of dark photons given by Eq.~(\ref{eq:emmissionrate}) one must thus distinguish transverse (T) and longitudinal (L) self-energy and so emission/absorption rates,  $( \Gamma, \Pi,) \rightarrow (\Gamma_{T,L}, \Pi_{T,L})$ \cite{An:2013yfc}. 

The transverse modes, corresponding to ordinary photons dressed by interaction with the medium, are the simplest. In the relativistic regime and to leading order in $\alpha$, the transverse photons simply get a thermal mass, albeit with a slight momentum dependence, 
\begin{equation}
\operatorname{Re}\Pi_{\gamma,T} = \left\{
\begin{array}{ll}
\omega_P^2 = \sum_i q_i^2 {T^2/9} &\mbox{low $k$}\\
{3/ 2}\, \omega_P^2 & \mbox{large $k$}
\end{array}\right.,
\end{equation}
where $\omega_P$ is the so-called plasma frequency and the sum is over relativistic charged particles \cite{Braaten:1993jw}. For later use, we mention the fact that there is no substantial wave-function renormalization for the transverse modes. 

\begin{center}
\begin{figure}[h!]
\includegraphics[scale=0.5]{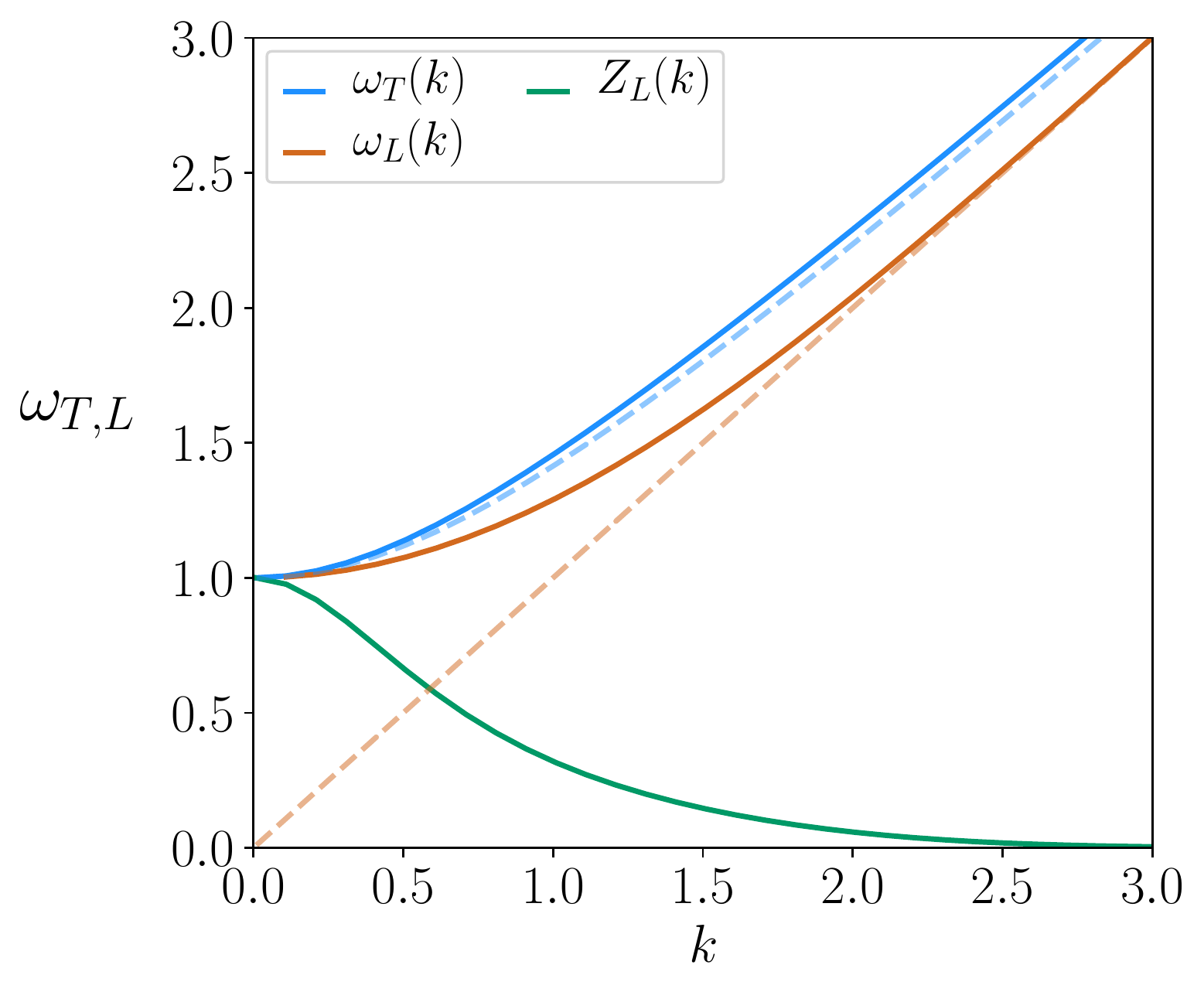}
\caption{Dispersion relations for $\omega_T(k)$ (blue solid) and $\omega_L(k)$ (orange solid) in the relativistic regime $T\ll m_e$. They are normalized to $\omega_P=1$. The dashed orange line is the dispersion relation of a standard massive particle. The dispersion relation of the transverse mode has $\omega_T= \omega_P$ for small $k$ and $\omega_T \approx \sqrt{3/2 \omega_P^2 + k^2}$ for large $k$. That of the longitudinal mode has $\omega_L = \omega_P$ for small $k$ but asymptots to $\omega_L \approx k$ for large $k$. Also shown is the wave-function normalization $Z_L$ (green solid) which goes to zero at large $k$, revealing that the longitudinal mode propagates only for small enough momenta. }\label{fig:omegaL}
\end{figure}
\end{center}

The longitudinal mode self-energy has a more involved structure. This is related to the fact that longitudinal photons do no propagate in vacuum. Still in the relativistic regime and to leading order in $\alpha$, the self-energy takes the form \cite{Braaten:1993jw}\footnote{We use here the definition of the longitudinal polarization tensor of  \cite{An:2013yfc,Redondo:2013lna}, $\Pi_L \equiv \Pi_L^{\rm APP}$ , which differs from that of \cite{Braaten:1993jw}, $\Pi_L^{\rm BS} = K^2/k^2 \Pi_L^{\rm APP}$.} 
\begin{equation}
\label{eq:PiL}
\Pi_{L}(\omega,k) = 3 \omega_P^2  {K^2\over k^2} \left({\omega\over 2 k} \log\left({\omega + k)\over (\omega - k)}\right) - 1\right).
\end{equation}
Solving for $ \omega_L^2 - k^2 = \operatorname{Re}\Pi_L(\omega_L(k), k)$ leads to
\begin{equation}
\operatorname{Re}\Pi_{\gamma,L} = \left\{
\begin{array}{ll}
\omega_P^2\,  K^2/\omega^2& k \sim 0\\
\sim 0 & k \gtrsim \omega_P
\end{array}\right. .
\end{equation}
The behaviors of the dispersion relations of the transverse (solid blue) and longitudinal (solid orange) modes are depicted in Fig.~\ref{fig:omegaL}. 
For a  longitudinal plasmon mode close to on-shell, $\omega \sim \omega_L$, $\operatorname{Re}\Pi_{\gamma,L}  \approx \omega_L^2\,  K^2/\omega^2$ and one can write \cite{Braaten:1993jw,An:2013yfc,Redondo:2013lna}\footnote{To be clear, while we consider the real part of the self-energy only to leading order in $\alpha$,the imaginary part $\operatorname{Im}\Pi_{L}$ may include higher order corrections in $\alpha$, so as to describe, for instance, 
Compton emission, etc.} 
\begin{equation}
\label{eq:plasmonOS}
 {1\over K^2 -\Pi_{L}}\approx{ \omega^2 \, Z_L\over K^2( \omega^2 - \omega_L^2) - i Z_L \omega^2 \operatorname{Im}\Pi_{L}},
\end{equation}
where, using Eq.~(\ref{eq:PiL}),  the wave-function normalization $Z_L$ is given by
\begin{equation}
Z_L^{-1} = 1+ {k^2\over \omega_L}  {3 \omega_P^2 - \omega_L^2 + k^2\over  2 (\omega_L^2 - k^2)},
\end{equation}
which satisfies  $Z_L \rightarrow 1$ as $k\rightarrow 0$ but goes to zero for $k\gtrsim \omega_P$, as shown  in Fig.~\ref{fig:omegaL} (green solid). This reflects the fact that the longitudinal mode mostly exists for moderate momenta. 
From this and (\ref{eq:emmissionrate}) we get\footnote{Expression (\ref{eq:GamLFinal})  agrees with the relevant literature \cite{An:2013yfc,Redondo:2013lna}, see also \cite{Rrapaj:2019eam} but perhaps a comment is in order. For instance, it may be compared with Eq.~(2.7) in \cite{Redondo:2013lna} using the fact that in Eq.~(\ref{eq:GamLFinal}) the longitudinal photon emission rates are to be calculated as in vacuum while in Ref.~\cite{Redondo:2013lna} (RR below) they include factors interpreted as  wave-function normalization: $\tilde Z_L \Gamma_{\rm \gamma,\, em}^L\vert_{\rm us}= \Gamma_{\rm \gamma,\, em}^L\vert_{\rm RR} $.}
\begin{equation}
\label{eq:GamLFinal}
\Gamma_{\rm \gamma',\, em}^L =   {\epsilon^2 m_{\gamma'}^4 \tilde Z_L^2 \Gamma_{\rm \gamma,\, em}^L  \over  (\omega^2 - \omega_L^2)^2 + \omega^2 (e^{\omega/T}-1)^2 (\tilde Z_L \Gamma_{\rm \gamma,\, em}^L)^2},
\end{equation}
with $\tilde Z_L = \omega^2/m_{\gamma'}^2\,  Z_L$. This is to be compared to 
\begin{equation}
\label{eq:GamRFinal}
\Gamma_{\rm \gamma',\, em}^T =   {\epsilon^2 m_{\gamma'}^4  \Gamma_{\rm \gamma,\, em}^T  \over  (m_{\gamma'}^2  - \omega_T^2)^2 + \omega^2 (e^{\omega/T}-1)^2 (\Gamma_{\rm \gamma,\, em}^T)^2}.
\end{equation}
Comparing these expressions, one sees that at large $T$, $\omega_P \gg m_{\gamma'}$ and $\omega \sim \omega_P$, 
\begin{equation}
{\Gamma_{\rm \gamma',\, em}^L \over \Gamma_{\rm \gamma',\, em}^T} \approx {\omega^4\over m_{\gamma'}^4} {\Gamma_{\rm \gamma,\, em}^L \over \Gamma_{\rm \gamma,\, em}^T} \label{eq::highT}.
\end{equation}
As the rate for production of longitudinal photon modes itself is $\propto m_{\gamma'}^2$, due to current conservation,
$k_\mu J^\mu = 0\rightarrow\epsilon_\mu^L J^\mu \propto m_{\gamma'}$, 
the overall scaling of this ratio is $\omega^2/m_{\gamma'}^2 \gg 1$. So production of dark photons is dominantly through production of longitudinal photons at high temperature, $\omega_P \gtrsim m_{\gamma'}$ \cite{An:2013yfc,Redondo:2013lna}. At lower temperatures however, which is the regime relevant for infrared dominated freeze-in production, 
\begin{equation}
{\Gamma_{\rm \gamma',\, em}^L \over \Gamma_{\rm \gamma',\, em}^T} \approx Z_L^2 {\Gamma_{\rm \gamma,\, em}^L \over \Gamma_{\rm \gamma,\, em}^T} \sim Z_L^2 {m_{\gamma'}\over \omega} \ll 1\label{eq::lowT},
\end{equation}
as $Z_L\gtrsim 1$ so production is dominantly through bona fide transverse photons \cite{An:2013yfc,Redondo:2013lna}. Thus, the production of on-shell dark photons proceeds essentially as in vacuum, but with cross-section in which the mixing parameter should be replaced by
\begin{equation}
\epsilon \rightarrow \epsilon_{\rm eff}^2 = {\epsilon^2 m_{\gamma'}^4 \over (m_{\gamma'}^2- \operatorname{Re}\Pi_{\gamma})^2 +  \omega^2(e^{\omega/T}-1)^2 \Gamma_{\rm \gamma}^2}
\end{equation}
legitimating the rule of thumb stated in Eq.~(\ref{eq:substitution}).\\

\begin{center}
\begin{figure}[h!]
\includegraphics[scale=0.5]{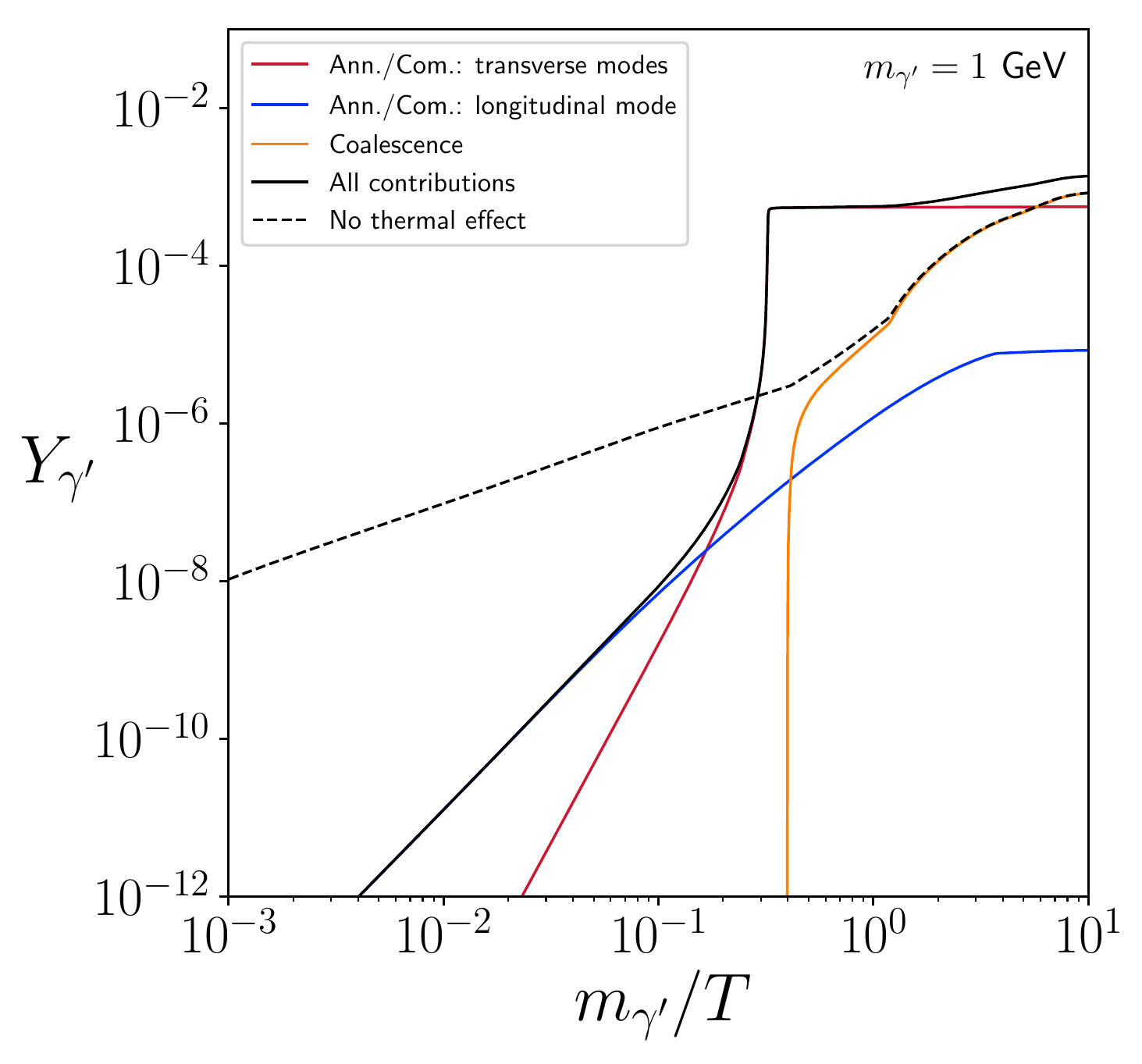}
\caption{For $\epsilon=10^{-9}$, all contributions to the dark photon yield as a function of the inverse temperature. One can distinguish contributions from pair annihilation and Compton processes for both transverses (red dashed) and longitudinal modes (blue dashed), from the coalescence process (orange dashed) and from all contributions together (solid black). For the coalescence, we have taken into account the thermal corrections to the mass of the SM particles that annihilate into a dark photon \cite{Redondo:2008ec}.}\label{fig::DPProductionChannels}
\end{figure}
\end{center}

Fig.~\ref{fig::DPProductionChannels} illustrates all the contributions to the production of dark photons. There, one recognises the high temperature suppression of transversal modes (Eq.~\ref{eq::highT}) and the low temperature suppression of the longitudinal mode (Eq.~\ref{eq::lowT}).

%%%%%%%%%%%%%%%%%%%%
\section{A note on dark Higgs production}
\label{sec:BEH}

\begin{figure}[ht]
\includegraphics[width=5cm]{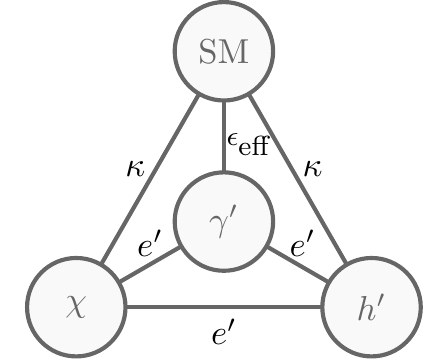}
\caption{The sectors and their connections in the Higgs phase of the dark photon model.}
\label{fig:Higgs}
\end{figure}

In the body of this work, we assumed that the only relevant degrees of freedom for DM production are, beside the SM particles, the dark photon and the dark matter itself. They thus apply as such if the mass of the dark photon arises through the  St\"uckelberg mechanism. If, instead, the $U(1)'$ is broken through the Brout-Englert-Higgs mechanism, our underlying basic assumption is that the dark Higgs ($h'$) is much heavier than both the dark photon and the dark matter, {\em i.e.} that
\begin{equation}
{m_{h'}\over m_{\gamma'}}= \sqrt{{\lambda'\over 2 \pi \alpha'}} \gg 1 \quad \mbox{\rm and}  \quad {m_{h'}\over m_{\rm DM}}  = {\sqrt{2 \lambda'} v'\over m_{\rm DM}}\gg 1,
\end{equation}
where $\lambda'$ and $v'$ are the dark Higgs  quartic coupling and vev. We assumed, for simplicity, that the DM is vector-like so that $m_{\rm DM}$ can be taken to be independent of $v'$, a hypothesis that could be relaxed in a more elaborated scenario. Now $\alpha'$ is a (very) tiny parameter for most of the parameter space we consider,  so these conditions can be easily met. If this is not the case (for instance if for some reason $\lambda'$ itself is a small parameter) then we may have {\em e.g.}
\begin{equation}
m_{h'} \sim m_{\gamma'} \ll m_{\rm DM}.
\end{equation}
If this is the case, the system becomes more complex, as  we must  in principle take into account the abundance of $h'$ particles on top of that of the DM and the dark photons,  see Fig.~\ref{fig:Higgs}. 

An interesting aspect of this scenario is that the process of associated production of dark Higgs (or dark-Higgsstrahlung) $\rm{SM} \rightarrow h' \gamma'$ is unsuppressed in the limit $m_{\gamma'}\rightarrow 0$, and scales as \cite{Pospelov:2008jk,Batell:2009yf} 
\begin{equation}
\sigma_{\rm SM \rightarrow h' \gamma'} \propto {\alpha^2 \kappa^2 \over s} 
\label{eq:HiggsS}
\end{equation}
for large $\sqrt{s} \gg (m_{h'} + m_{\gamma'})$. As $\kappa = \epsilon \sqrt{\alpha'/\alpha}$, the rate of $h'$ production is proportional to $\alpha'$ and not $\alpha'^2$ as naively expected from the vertex ${\cal L} \supset e'^2 v' h' A_\mu' A'^\mu \equiv e' m_{\gamma'} h' A_\mu' A'^\mu$. 
This result, familiar from SM Higgs physics, can be traced back to the fact that the process (\ref{eq:HiggsS})  is dominated by longitudinal $\gamma'$ emission at large energies. Hence, the connection between the SM sector and the $h'$ is controlled by $\kappa$, precisely like the DM itself, see Fig.~\ref{fig:Higgs}. By the same token, the connections between the $h'$, the $\gamma'$ and the DM are all driven by $e'$. 

From the above discussion and the (approximate) symmetry of Fig.~\ref{fig:Higgs} we thus expect that the production of the dark Higgs will proceed as that of DM, starting from the freeze-in regime  Ia for tiny values of $\alpha'$, then Ib with production of $h'$ and DM from $\gamma'$ in thermal equilibrium with the SM sector, etc. So the abundance of $h'$ should essentially track that of DM, $Y_{h'} \sim Y_{\rm DM}$, and this at least as long as $T \gtrsim m_{\rm DM}$. Thus, if sequential freeze-in occurs, the contribution of the process $h' \gamma' \rightarrow \chi \bar \chi$ should be subdominant compared to that of the process $\gamma' \gamma' \rightarrow \chi \bar \chi$. Consequently, and while a more detailed analysis may be of interest, we  tentatively conclude that, generically speaking,  the presence of $h'$ {should} not affect significantly the structure of the phase diagram depicted in Figs.~\ref{fig::MESA_3_1} and \ref{fig::MESA_100_10}. 

As for the symmetric phase, in which $m_{\gamma'} = 0$, it would contain two DM candidates, as the complex scalar associated to the dark Higgs would also be stable (by charge conservation), and both would be produced at similar rates for tiny couplings. In particular, as the abundance at freeze-in is $Y_{\rm DM} \propto 1/m_{\rm DM}$ (baring production from Z boson decay, see \cite{Chu:2011be}), both particles will give (roughly) a similar contribution to the DM energy density, $\Omega_{\rm DM}\propto m_{\rm DM} Y_{\rm DM}$. While reannihilation and subsequent regimes may {have a richer structure than} in a scheme with a single DM candidate, we expect the phase diagram to follow essentially the "mesa" pattern studied in \cite{Chu:2011be}. 

\bibliographystyle{JHEP}
\bibliography{biblio}

\providecommand{\href}[2]{#2}\begingroup\raggedright\begin{thebibliography}{10}

\bibitem{Patt:2006fw}
B.~Patt and F.~Wilczek, \emph{{Higgs-field portal into hidden sectors}},
  \href{http://arxiv.org/abs/hep-ph/0605188}{{\tt hep-ph/0605188}}.

\bibitem{Chu:2011be}
X.~Chu, T.~Hambye and M.~H.~G. Tytgat, \emph{{The Four Basic Ways of Creating
  Dark Matter Through a Portal}},
  \href{http://dx.doi.org/10.1088/1475-7516/2012/05/034}{\emph{JCAP} {\bf 1205}
  (2012) 034}, [\href{http://arxiv.org/abs/1112.0493}{{\tt 1112.0493}}].

\bibitem{Jaeckel:2010ni}
J.~Jaeckel and A.~Ringwald, \emph{{The Low-Energy Frontier of Particle
  Physics}},
  \href{http://dx.doi.org/10.1146/annurev.nucl.012809.104433}{\emph{Ann. Rev.
  Nucl. Part. Sci.} {\bf 60} (2010) 405--437},
  [\href{http://arxiv.org/abs/1002.0329}{{\tt 1002.0329}}].

\bibitem{Essig:2013lka}
R.~Essig et~al., \emph{{Working Group Report: New Light Weakly Coupled
  Particles}},  in \emph{{Proceedings, 2013 Community Summer Study on the
  Future of U.S. Particle Physics: Snowmass on the Mississippi (CSS2013):
  Minneapolis, MN, USA, July 29-August 6, 2013}}, 2013.
\newblock \href{http://arxiv.org/abs/1311.0029}{{\tt 1311.0029}}.

\bibitem{Alexander:2016aln}
J.~Alexander et~al., \emph{{Dark Sectors 2016 Workshop: Community Report}},
  2016.
\newblock \href{http://arxiv.org/abs/1608.08632}{{\tt 1608.08632}}.

\bibitem{Foot:2014uba}
R.~Foot and S.~Vagnozzi, \emph{{Dissipative hidden sector dark matter}},
  \href{http://dx.doi.org/10.1103/PhysRevD.91.023512}{\emph{Phys. Rev.} {\bf
  D91} (2015) 023512}, [\href{http://arxiv.org/abs/1409.7174}{{\tt
  1409.7174}}].

\bibitem{Ackerman:mha}
L.~Ackerman, M.~R. Buckley, S.~M. Carroll and M.~Kamionkowski, \emph{{Dark
  Matter and Dark Radiation}},
  \href{http://dx.doi.org/10.1103/PhysRevD.79.023519,
  10.1142/9789814293792_0021}{\emph{Phys. Rev.} {\bf D79} (2009) 023519},
  [\href{http://arxiv.org/abs/0810.5126}{{\tt 0810.5126}}].

\bibitem{Feng:2008mu}
J.~L. Feng, H.~Tu and H.-B. Yu, \emph{{Thermal Relics in Hidden Sectors}},
  \href{http://dx.doi.org/10.1088/1475-7516/2008/10/043}{\emph{JCAP} {\bf 0810}
  (2008) 043}, [\href{http://arxiv.org/abs/0808.2318}{{\tt 0808.2318}}].

\bibitem{Feng:2009mn}
J.~L. Feng, M.~Kaplinghat, H.~Tu and H.-B. Yu, \emph{{Hidden Charged Dark
  Matter}}, \href{http://dx.doi.org/10.1088/1475-7516/2009/07/004}{\emph{JCAP}
  {\bf 0907} (2009) 004}, [\href{http://arxiv.org/abs/0905.3039}{{\tt
  0905.3039}}].

\bibitem{Hambye:2010zb}
T.~Hambye, \emph{{On the stability of particle dark matter}},
  \href{http://dx.doi.org/10.22323/1.110.0098}{\emph{PoS} {\bf IDM2010} (2011)
  098}, [\href{http://arxiv.org/abs/1012.4587}{{\tt 1012.4587}}].

\bibitem{Holdom:1985ag}
B.~Holdom, \emph{{Two U(1)'s and Epsilon Charge Shifts}},
  \href{http://dx.doi.org/10.1016/0370-2693(86)91377-8}{\emph{Phys. Lett.} {\bf
  166B} (1986) 196--198}.

\bibitem{Stueckelberg:1900zz}
E.~C.~G. Stueckelberg, \emph{{Interaction energy in electrodynamics and in the
  field theory of nuclear forces}},
  \href{http://dx.doi.org/10.5169/seals-110852}{\emph{Helv. Phys. Acta} {\bf
  11} (1938) 225--244}.

\bibitem{Englert:1964et}
F.~Englert and R.~Brout, \emph{{Broken Symmetry and the Mass of Gauge Vector
  Mesons}}, \href{http://dx.doi.org/10.1103/PhysRevLett.13.321}{\emph{Phys.
  Rev. Lett.} {\bf 13} (1964) 321--323}.

\bibitem{Higgs:1964pj}
P.~W. Higgs, \emph{{Broken Symmetries and the Masses of Gauge Bosons}},
  \href{http://dx.doi.org/10.1103/PhysRevLett.13.508}{\emph{Phys. Rev. Lett.}
  {\bf 13} (1964) 508--509}.

\bibitem{McDonald:2001vt}
J.~McDonald, \emph{{Thermally generated gauge singlet scalars as
  selfinteracting dark matter}},
  \href{http://dx.doi.org/10.1103/PhysRevLett.88.091304}{\emph{Phys. Rev.
  Lett.} {\bf 88} (2002) 091304},
  [\href{http://arxiv.org/abs/hep-ph/0106249}{{\tt hep-ph/0106249}}].

\bibitem{Hall:2009bx}
L.~J. Hall, K.~Jedamzik, J.~March-Russell and S.~M. West, \emph{{Freeze-In
  Production of FIMP Dark Matter}},
  \href{http://dx.doi.org/10.1007/JHEP03(2010)080}{\emph{JHEP} {\bf 03} (2010)
  080}, [\href{http://arxiv.org/abs/0911.1120}{{\tt 0911.1120}}].

\bibitem{Bernal:2017kxu}
N.~Bernal, M.~Heikinheimo, T.~Tenkanen, K.~Tuominen and V.~Vaskonen, \emph{{The
  Dawn of FIMP Dark Matter: A Review of Models and Constraints}},
  \href{http://dx.doi.org/10.1142/S0217751X1730023X}{\emph{Int. J. Mod. Phys.}
  {\bf A32} (2017) 1730023}, [\href{http://arxiv.org/abs/1706.07442}{{\tt
  1706.07442}}].

\bibitem{Pospelov:2007mp}
M.~Pospelov, A.~Ritz and M.~B. Voloshin, \emph{{Secluded WIMP Dark Matter}},
  \href{http://dx.doi.org/10.1016/j.physletb.2008.02.052}{\emph{Phys. Lett.}
  {\bf B662} (2008) 53--61}, [\href{http://arxiv.org/abs/0711.4866}{{\tt
  0711.4866}}].

\bibitem{Pospelov:2008zw}
M.~Pospelov, \emph{{Secluded U(1) below the weak scale}},
  \href{http://dx.doi.org/10.1103/PhysRevD.80.095002}{\emph{Phys. Rev.} {\bf
  D80} (2009) 095002}, [\href{http://arxiv.org/abs/0811.1030}{{\tt
  0811.1030}}].

\bibitem{Foot:1994bx}
R.~Foot, \emph{{Experimental signatures of a massive mirror photon}},
  \href{http://arxiv.org/abs/hep-ph/9407331}{{\tt hep-ph/9407331}}.

\bibitem{Gondolo:1990dk}
P.~Gondolo and G.~Gelmini, \emph{{Cosmic abundances of stable particles:
  Improved analysis}},
  \href{http://dx.doi.org/10.1016/0550-3213(91)90438-4}{\emph{Nucl. Phys.} {\bf
  B360} (1991) 145--179}.

\bibitem{Kane:2015qea}
G.~L. Kane, P.~Kumar, B.~D. Nelson and B.~Zheng, \emph{{Dark matter production
  mechanisms with a nonthermal cosmological history: A classification}},
  \href{http://dx.doi.org/10.1103/PhysRevD.93.063527}{\emph{Phys. Rev.} {\bf
  D93} (2016) 063527}, [\href{http://arxiv.org/abs/1502.05406}{{\tt
  1502.05406}}].

\bibitem{Klasen:2013ypa}
M.~Klasen and C.~E. Yaguna, \emph{{Warm and cold fermionic dark matter via
  freeze-in}},
  \href{http://dx.doi.org/10.1088/1475-7516/2013/11/039}{\emph{JCAP} {\bf 1311}
  (2013) 039}, [\href{http://arxiv.org/abs/1309.2777}{{\tt 1309.2777}}].

\bibitem{Redondo:2008aa}
J.~Redondo, \emph{{Helioscope Bounds on Hidden Sector Photons}},
  \href{http://dx.doi.org/10.1088/1475-7516/2008/07/008}{\emph{JCAP} {\bf 0807}
  (2008) 008}, [\href{http://arxiv.org/abs/0801.1527}{{\tt 0801.1527}}].

\bibitem{Jaeckel:2008fi}
J.~Jaeckel, J.~Redondo and A.~Ringwald, \emph{{Signatures of a hidden cosmic
  microwave background}},
  \href{http://dx.doi.org/10.1103/PhysRevLett.101.131801}{\emph{Phys. Rev.
  Lett.} {\bf 101} (2008) 131801}, [\href{http://arxiv.org/abs/0804.4157}{{\tt
  0804.4157}}].

\bibitem{Redondo:2008ec}
J.~Redondo and M.~Postma, \emph{{Massive hidden photons as lukewarm dark
  matter}}, \href{http://dx.doi.org/10.1088/1475-7516/2009/02/005}{\emph{JCAP}
  {\bf 0902} (2009) 005}, [\href{http://arxiv.org/abs/0811.0326}{{\tt
  0811.0326}}].

\bibitem{An:2013yfc}
H.~An, M.~Pospelov and J.~Pradler, \emph{{New stellar constraints on dark
  photons}},
  \href{http://dx.doi.org/10.1016/j.physletb.2013.07.008}{\emph{Phys. Lett.}
  {\bf B725} (2013) 190--195}, [\href{http://arxiv.org/abs/1302.3884}{{\tt
  1302.3884}}].

\bibitem{Redondo:2013lna}
J.~Redondo and G.~Raffelt, \emph{{Solar constraints on hidden photons
  re-visited}},
  \href{http://dx.doi.org/10.1088/1475-7516/2013/08/034}{\emph{JCAP} {\bf 1308}
  (2013) 034}, [\href{http://arxiv.org/abs/1305.2920}{{\tt 1305.2920}}].

\bibitem{Fradette:2014sza}
A.~Fradette, M.~Pospelov, J.~Pradler and A.~Ritz, \emph{{Cosmological
  Constraints on Very Dark Photons}},
  \href{http://dx.doi.org/10.1103/PhysRevD.90.035022}{\emph{Phys. Rev.} {\bf
  D90} (2014) 035022}, [\href{http://arxiv.org/abs/1407.0993}{{\tt
  1407.0993}}].

\bibitem{Weldon:1983jn}
H.~A. Weldon, \emph{{Simple Rules for Discontinuities in Finite Temperature
  Field Theory}}, \href{http://dx.doi.org/10.1103/PhysRevD.28.2007}{\emph{Phys.
  Rev.} {\bf D28} (1983) 2007}.

\bibitem{Bellac:2011kqa}
M.~L. Bellac, \emph{{Thermal Field Theory}}.
\newblock Cambridge Monographs on Mathematical Physics. Cambridge University
  Press, 2011.
\newblock 10.1017/CBO9780511721700.

\bibitem{Braaten:1993jw}
E.~Braaten and D.~Segel, \emph{{Neutrino energy loss from the plasma process at
  all temperatures and densities}},
  \href{http://dx.doi.org/10.1103/PhysRevD.48.1478}{\emph{Phys. Rev.} {\bf D48}
  (1993) 1478--1491}, [\href{http://arxiv.org/abs/hep-ph/9302213}{{\tt
  hep-ph/9302213}}].

\bibitem{Hambye:2018dpi}
T.~Hambye, M.~H.~G. Tytgat, J.~Vandecasteele and L.~Vanderheyden, \emph{{Dark
  matter direct detection is testing freeze-in}},
  \href{http://dx.doi.org/10.1103/PhysRevD.98.075017}{\emph{Phys. Rev.} {\bf
  D98} (2018) 075017}, [\href{http://arxiv.org/abs/1807.05022}{{\tt
  1807.05022}}].

\bibitem{Essig:2011nj}
R.~Essig, J.~Mardon and T.~Volansky, \emph{{Direct Detection of Sub-GeV Dark
  Matter}}, \href{http://dx.doi.org/10.1103/PhysRevD.85.076007}{\emph{Phys.
  Rev.} {\bf D85} (2012) 076007}, [\href{http://arxiv.org/abs/1108.5383}{{\tt
  1108.5383}}].

\bibitem{wip}
T.~Hambye, M.~H. Tytgat, J.~Vandecasteele and L.~Vanderheyden, \emph{{work in
  progress}}, .

\bibitem{Berger:2016vxi}
J.~Berger, K.~Jedamzik and D.~G.~E. Walker, \emph{{Cosmological Constraints on
  Decoupled Dark Photons and Dark Higgs}},
  \href{http://dx.doi.org/10.1088/1475-7516/2016/11/032}{\emph{JCAP} {\bf 1611}
  (2016) 032}, [\href{http://arxiv.org/abs/1605.07195}{{\tt 1605.07195}}].

\bibitem{Weinberg:1979bt}
S.~Weinberg, \emph{{Cosmological Production of Baryons}},
  \href{http://dx.doi.org/10.1103/PhysRevLett.42.850}{\emph{Phys. Rev. Lett.}
  {\bf 42} (1979) 850--853}.

\bibitem{Rrapaj:2019eam}
E.~Rrapaj, A.~Sieverding and Y.-Z. Qian, \emph{{Rate of dark photon emission
  from electron positron annihilation in massive stars}},
  \href{http://arxiv.org/abs/1904.10567}{{\tt 1904.10567}}.

\bibitem{Pospelov:2008jk}
M.~Pospelov, A.~Ritz and M.~B. Voloshin, \emph{{Bosonic super-WIMPs as
  keV-scale dark matter}},
  \href{http://dx.doi.org/10.1103/PhysRevD.78.115012}{\emph{Phys. Rev.} {\bf
  D78} (2008) 115012}, [\href{http://arxiv.org/abs/0807.3279}{{\tt
  0807.3279}}].

\bibitem{Batell:2009yf}
B.~Batell, M.~Pospelov and A.~Ritz, \emph{{Probing a Secluded U(1) at
  B-factories}},
  \href{http://dx.doi.org/10.1103/PhysRevD.79.115008}{\emph{Phys. Rev.} {\bf
  D79} (2009) 115008}, [\href{http://arxiv.org/abs/0903.0363}{{\tt
  0903.0363}}].

\end{thebibliography}\endgroup

\end{document}